\documentclass[reprint,superscriptaddress,nofootinbib,longbibliography]{revtex4-2}

\usepackage{eqnarray,physics,bbold,dsfont}
\usepackage{trackchanges}
\usepackage{graphicx}
\usepackage{breqn}
\usepackage[dvipsnames]{xcolor}
\usepackage{appendix}
\usepackage{lineno}

\DeclareMathAlphabet{\mathpzc}{OT1}{pzc}{m}{it}

\newcommand{\vek}{\Vec{k}}

 \newcommand{\nmean}[1]{\langle n_{#1} \rangle}

\title{Effect of confinement and Coulomb interactions on the electronic structure of (111) LaAlO$_3$/SrTiO$_3$ interface}

\newcommand{\INFN}{INFN - Sezione di Napoli, Complesso Univ. Monte S. Angelo, I-80126 Napoli, Italy}
\newcommand{\UNINA}{Physics Department "Ettore Pancini", Universit\'a degli studi di Napoli "Federico II", Complesso Univ. Monte S. Angelo, Via Cintia, I-80126 Napoli, Italy}
\newcommand{\UNISA}{Physics Department "E.R. Caianiello", Universit\'a degli studi di Salerno, Via Giovanni Paolo II, 132, I-84084 Fisciano (Sa), Italy}
\newcommand{\CNR}{CNR-SPIN Napoli Unit, Complesso Univ. Monte S. Angelo, Via Cintia, I-80126 Napoli, Italy}
\newcommand{\IFW}{Institute for Theoretical Solid State Physics, IFW Dresden, Helmholtzstr. 20, 01069 Dresden, Germany}

\begin{document}
\title{Effect of confinement and Coulomb interactions on the electronic structure of (111) LaAlO$_3$/SrTiO$_3$ interface}

\author{M. Trama}
\email{mtrama@unisa.it}
\affiliation{\UNISA}
\affiliation{\INFN}
\affiliation{\IFW}

\author{V. Cataudella}
\affiliation{\UNINA}
\affiliation{\CNR}

\author{C. A. Perroni}
\affiliation{\UNINA}
\affiliation{\CNR}

\author{F. Romeo}
\affiliation{\UNISA}

\author{R. Citro}
\email{rocitro@unisa.it}
\affiliation{\UNISA}
\affiliation{\INFN}

\begin{abstract}
    A tight binding supercell approach is used for the calculation of the electronic structure of the (111) LaAlO$_3$/SrTiO$_3$ interface. The confinement potential at the interface is evaluated solving a discrete Poisson equation by means of an iterative method. In addition to the effect of the confinement, local Hubbard electron-electron terms are included at mean-field level within a fully self-consistent procedure. The calculation carefully describes how the two-dimensional electron gas arises from the quantum confinement of electrons near the interface due to band bending potential. The resulting electronic sub-bands and Fermi surfaces show full agreement with the electronic structure determined by angle-resolved photoelectron spectroscopy experiments. In particular, it is analyzed how the effect of local Hubbard interactions changes the density distribution over the layers from the interface to the bulk. Interestingly, the two-dimensional electron gas at interface is not depleted by local Hubbard interactions which indeed induce an enhancement of the electron density between the first layers and the bulk.
\end{abstract}

\maketitle
\onecolumngrid
\section{Introduction}
Recently, the emergent field of oxide electronics has revealed a rich phenomenology connected to the
creation and manipulation of interface electronic states.
After the discovery of two-dimensional electron gas (2DEG) at the (001) interface between the perovskite band insulators SrTiO$_3$ (STO) and LaAlO$_3$(LAO)~\cite{ohtomo2004high}, which are characterized by high-mobility, much work has been devoted to revealing its properties, like gate-controlled metal-insulator transitions~\cite{caviglia2008electric}, superconductivity~\cite{reyren2007superconducting}, including topological one~\cite{perroni1,maiellaro2019unveiling}, and its possible coexistence with magnetism~\cite{barthelemy2021quasi}. 
Recently, the successful creation of 2DEGs at the (111)-oriented interface of LAO/STO~\cite{davis2018anisotropic,monteiro2019band} has opened the possibility to investigate intriguing phenomena related to topological phase transitions~\cite{trama2021straininduced}, gate tunable anomaluos Hall effect~\cite{trama2022gate} and the spin/orbital Edelstein effect~\cite{trama2022tunable}. Despite the great number of works a supplement of analysis of the band structure of (111) LAO/STO is still required, especially in relation to confinement effects and the role of electronic correlations. \\ 
A first qualitative understanding of the band structure and Fermi surface has come from a tight-binding (TB) supercell calculation based on an ab initio bulk band structure discussed in~\cite{walker2014control}. The calculation of the surface electronic structure was performed by introducing a supercell containing 120 Ti atoms stacked along the $(111)$ direction and using maximally localized Wannier functions with additional on-site potential terms to account for band bending via an electrostatic potential. The TB Hamiltonian was solved self-consistently with Poisson's equation, incorporating an electric field dependent dielectric constant~\cite{king2014quasiparticle,bahramy2012emergent} and with only an adjustable parameter, the total magnitude of the band bending at the surface~\cite{walker2014control}. The derived Fermi surface (FS) consists of three equivalent elliptical sheets oriented along $\Gamma-M$ direction. The band structure along $\Gamma-M$ direction shows a single heavy band, corresponding to the long axis of one of the FS ellipses, which is nearly degenerate at the band bottom with a more dispersive, doubly degenerate band arising from the two remaining FS sheets. The band structure shows three confined 2DEG subbands arising from the $t_{2g}$ orbitals and a “ladder” of states with a bulk-like character above Fermi level due to the finite size of the supercell. The second subband was predicted to be just below the Fermi energy, in good agreement with the angle-resolved photoemission spectroscopy (ARPES) experiments. The wave functions of the lowest subband at the $\Gamma$ point was predicted to be extended over $\simeq 15$ Ti layers, an order of magnitude more than the lowest bulk subband on (001) STO, due to the lighter effective masses. \\
In this work we perform a TB supercell calculation for (111) LAO/STO, and crucially, beyond the effect of the confinement, we also include local Hubbard electron-electron interactions within a fully self-consistent mean-field approach. Furthermore, compared with Ref.~\cite{walker2014control}, we also account for the impact of SOC. In particular, the TB supercell Hamiltonian in the (111) direction is obtained by rotating the coordinates and converting the quasi-momentum degree of freedom along the (111) direction to the discrete index numbering the layer of Ti along the (111) axis.
Our calculation shows full agreement with the observed electronic structure by ARPES~\cite{walker2014control} and describes how the 2DEG arises from the quantum confinement of $t_{2g}$ electrons near the surface due to band bending. Moreover, we also demonstrate how the effect of local Hubbard terms changes the density distribution over the layers close to the surface. We show that, contrary to a naif expectation, the 2DEG at interface is not depleted by local Hubbard interactions which instead induce a modulation of the electron density as a function of the layer number. In fact, we find that local Hubbard terms enhance the electron density between the first layers of the interface and the bulk. \\
The manuscript is organized as follows. In Sec.~\ref{sec:method} we introduce the Hamiltonian, the TB supercell approach and we present the results for band structure, Fermi surface, and self-consistent band bending potential. We analyze the effects  both the absence and the presence of local Hubbard electron-electron interactions. In Sec.~\ref{sec:discussion} we discuss our results and we give a comparative discussion of previous studies. 


\section{Methods and results}
\label{sec:method}
In this Section we present the model and the results obtained within a TB supercell approach, both in absence and in presence of local Hubbard electron-electron interactions.

\subsection{Model}
\begin{figure}
    \centering
    \includegraphics[width=0.93\textwidth]{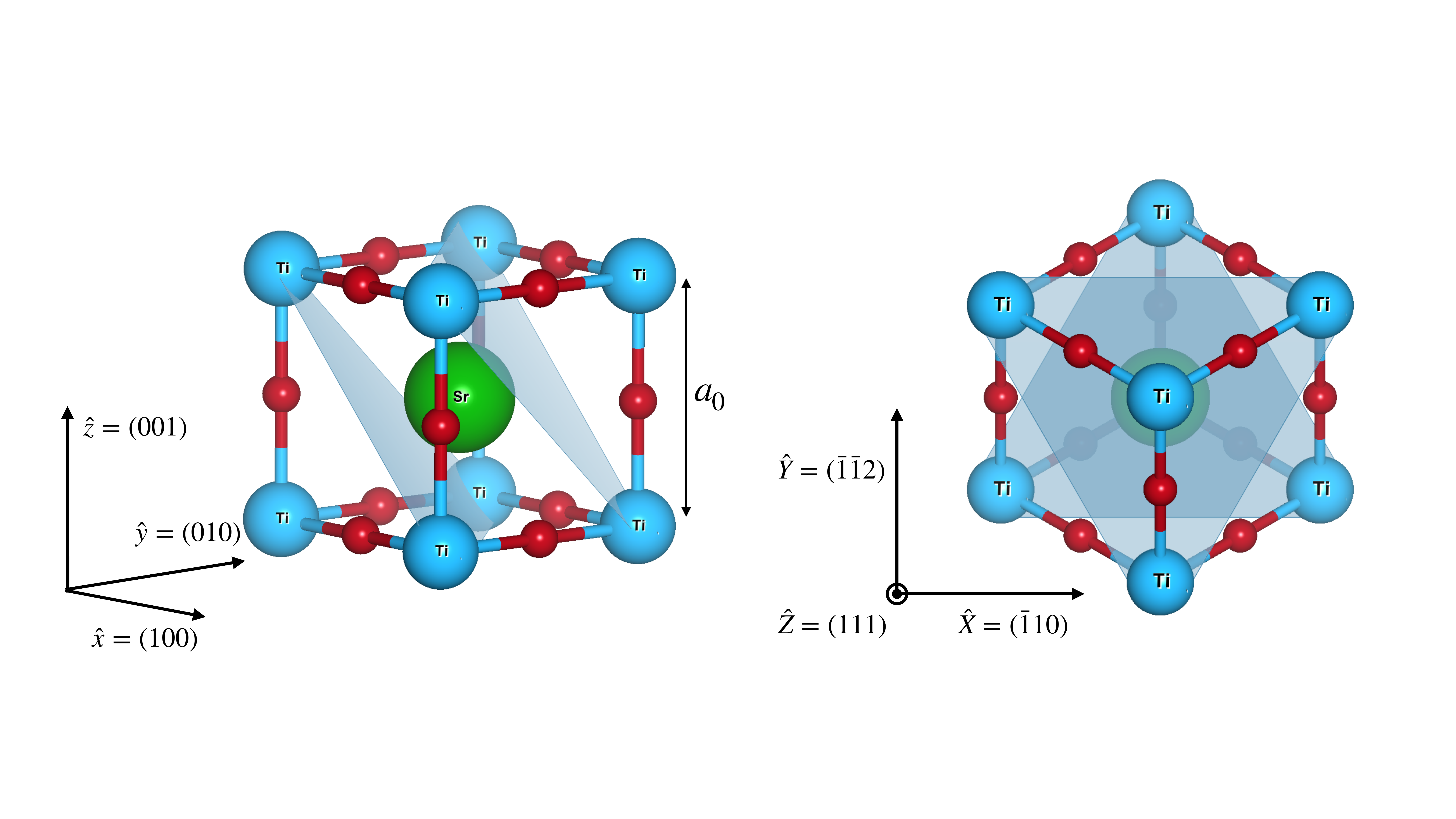}
    \caption{(Left panel) Crystal structure of SrTiO$_3$. The structure is cubic with a lattice parameter of $a_0=0.3905$ nm. The blue dots represents the Ti atoms, the red dots the oxygen atoms, and the green dot the Sr atom. (Right panel) The crystal structure seen from the (111) direction. The projection of Ti atoms along the (111) plane forms a honeycomb lattice. The figure has been generated through Vesta software~\cite{Momma11Vesta}.}
    \label{fig:Structure}
\end{figure}
STO has a cubic perovskite structure as shown in Fig.~\ref{fig:Structure}. The conductance bands form out of the $t_{2g}=\{d_{yz},d_{zx},d_{xy}\}$ orbitals of the Ti atoms in the bulk structure. Therefore, we focus only on the Ti lattice, which has a simple cubic structure at room temperature~\cite{pai2018physics} with a lattice constant $a_0=0.3905$ nm. The presence of a thin film of LAO over STO leads to the formation of a 2DEG and a thin positive charge density at its top. This is mostly ascribed to oxygen vacancies in STO~\cite{walker2014control} leading to an electronic reconfiguration which neutralizes this positive charge. As a consequence, the conduction band is partially filled, so that the electronic properties are determined by the low energy region of such bands.
In the present work we will take the positive charge at the interface as a free parameter of the model. In this sense we manipulate the number of oxygen vacancies.
\\In order to reconstruct the self-consistent electronic band structure, we adopt a TB Hamiltonian framed in the basis of atomic orbitals, using hopping parameters which fit the available ARPES data for the lowest bands~\cite{walker2014control}.
The bulk Hamiltonian for the conductance bands in STO, expressed in the quasi-momentum $(K_x,K_y,K_z)$ directed along the cubic axes, is
\begin{equation}
    H_{\text{TB}}^{\text{Bulk}}=\sum_{\vec{K}}\sum_{i\neq j\neq k}\sum_{\sigma} \left(-t_D\cos(K_i a_0)-t_D\cos(K_j a_0)-t_I\cos(K_k a_0))\right) d_{ij,\sigma,\vec{K}}^\dagger d_{ij,\sigma,\vec{K}}
    \label{eq:bulk_ham}
\end{equation}
where we truncated to nearest-neighbour hopping. 
Here $\{i,j,k\}$ runs over $\{x,y,z\}$, $d_{ij,\sigma,\vec{K}}$ is the annihilation operator of the electron characterized by $d_{ij}$ orbital, spin $\sigma$ and quasi-momentum $\vec{K}$. $t_D$ and $t_I$ are the direct and indirect hopping parameters, which we choose to be $t_D=0.25$ eV and $t_I=0.02$ eV~\cite{trama2021straininduced} in agreement with ARPES data.
Since the electric field produced from the interfacial charge breaks translational invariance along the (111) direction, the Hamiltonian expressed in terms of the quasi-momentum component along (111) is not the optimal choice for the description of the two-dimensional gas. Therefore,
from this Hamiltonian, we construct a TB supercell Hamiltonian in the (111) direction by a rotation of coordinates and converting the quasi-momentum degree of freedom along the (111) direction to the discrete index numbering the layer of Ti along the (111) axis. By this procedure, we convert the $6\times 6$ bulk Hamiltonian (considering the spin degree of freedom) to a $6N\times6N$ Hamiltonian, for which $N$ represents the number of layers considered (in the bulk system $N\to\infty$).
We include two other local terms in real coordinates, which are therefore independent of $\vec{K}$: the spin-orbit coupling (SOC) $H_{\text{SOC}}$, and a trigonal crystal field along the (111) direction $H_{\text{TRI}}$~\cite{trama2021straininduced,trama2022gate,trama2022tunable,xiao2011interface}.
The matrix for the TB supercell Hamiltonian has the form
\begin{equation}
    H=\begin{pmatrix}
        H_0 & H_t & 0 & 0 & 0 &...\\
        H_t^{\dagger} & H_0 & H_t & 0 & 0 & ...\\
        0 & H_t^{\dagger} & H_0 & H_t & 0 & ...\\
        0 & 0 & H_t^{\dagger} & H_0 & H_t & ...\\
        ... & ... & ... & ... & ... & ...\\
    \end{pmatrix},
    \label{eq:Ham_form}
\end{equation}
where 
\begin{equation}
    H_0=H_{\text{SOC}}+H_{\text{TRI}}
\end{equation}
and $H_t$ is the tunneling Hamiltonian describing the hopping between two neighboring layers for a given state of defined quasi-momentum parallel to the interface.
\begin{figure}
    \centering
    \includegraphics[width=0.63\textwidth]{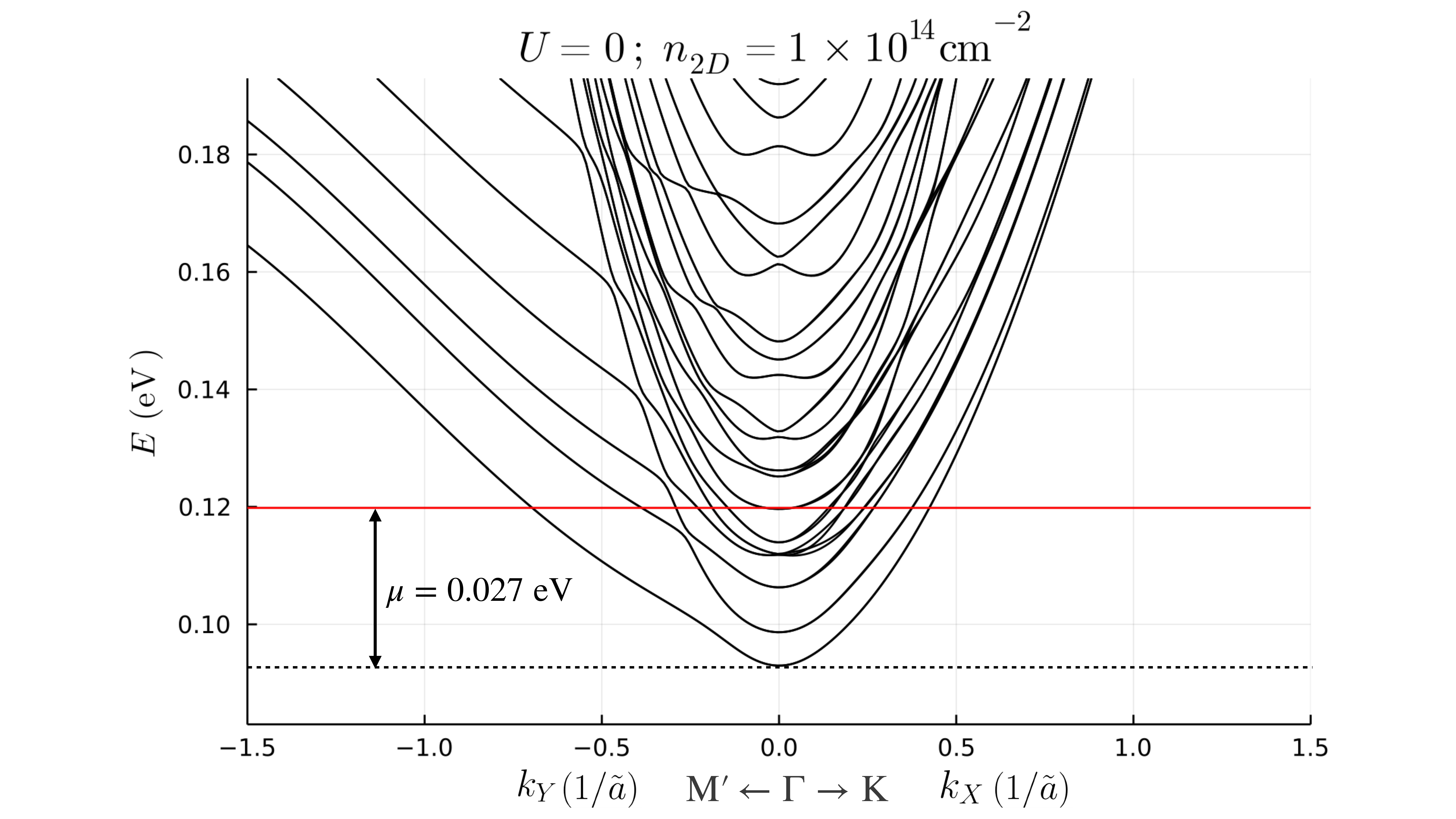}\\
    \includegraphics[width=0.63\textwidth]{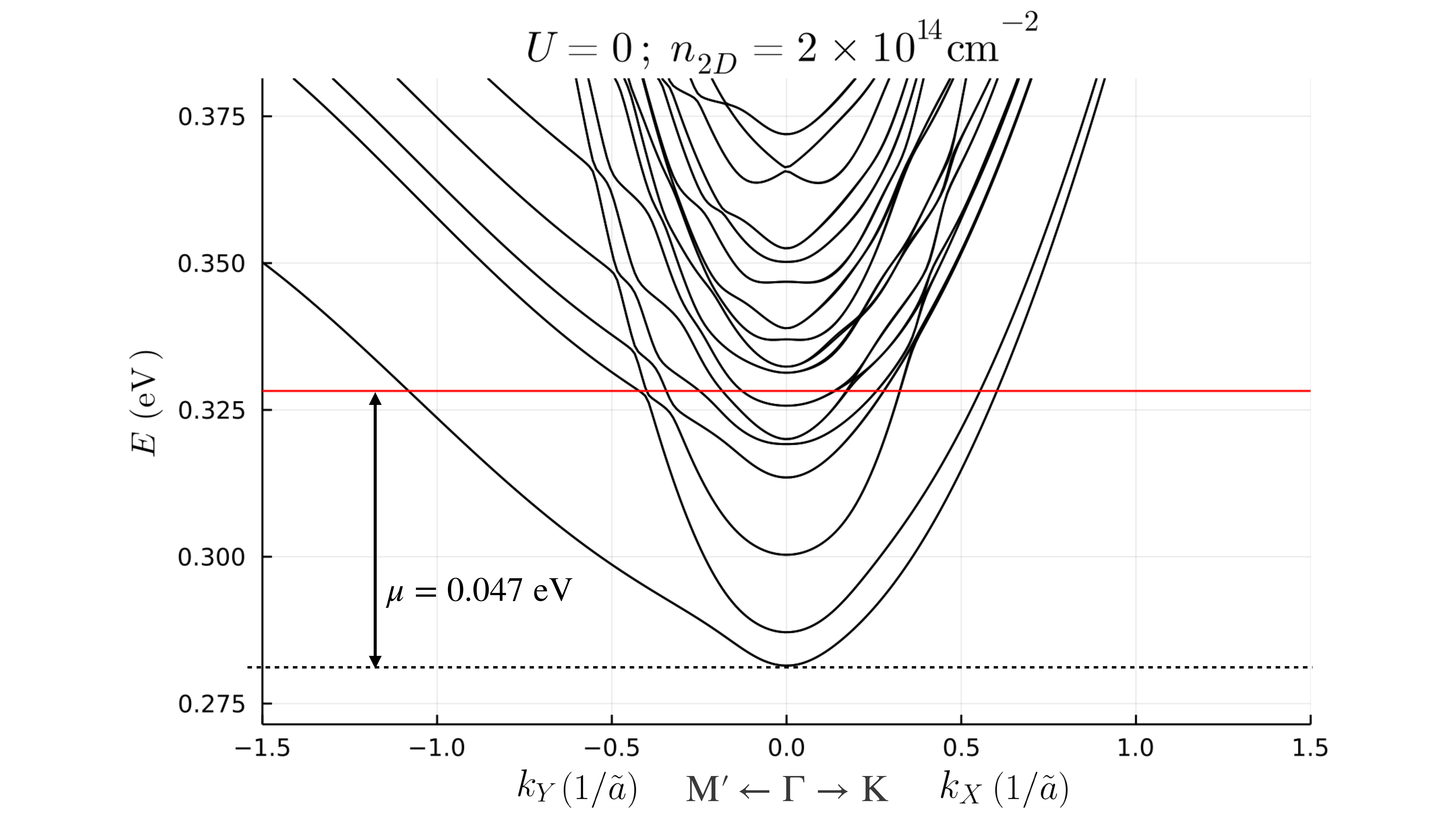}\\
    \includegraphics[width=0.63\textwidth]{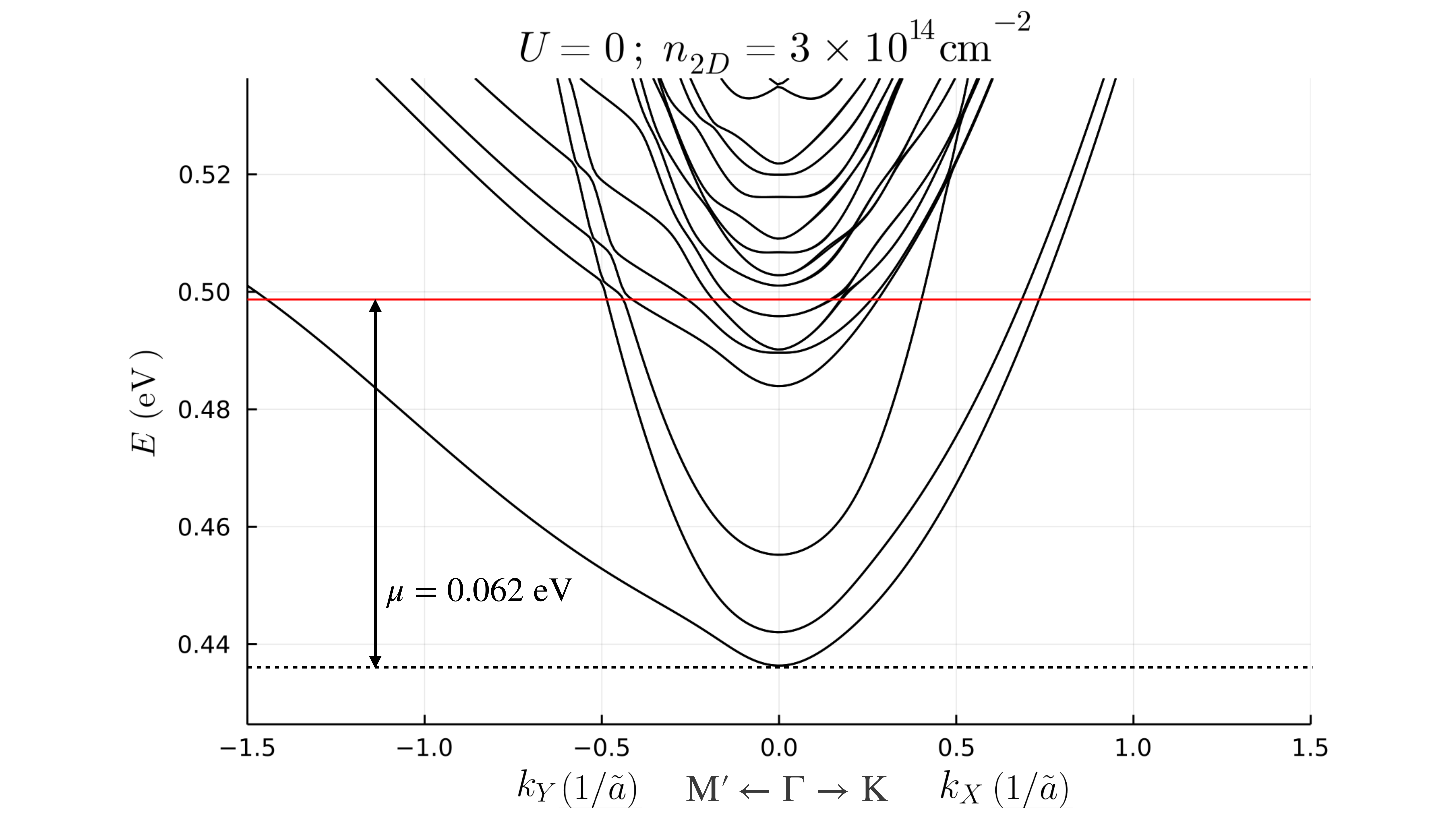}
    \caption{Low filling electronic band structure for the benchmark choice of density $n_{2D}=1\times10^{14}$ cm$^{-2}$ (upper panel), $n_{2D}=1\times10^{14}$ cm$^{-2}$ (middle panel) and $n_{2D}=3\times10^{14}$ cm$^{-2}$ (lower panel). The red lines is the Fermi level of the system.}
    \label{fig:bands_U0}
\end{figure}
In the Appendix~\ref{AppendiceA} we give all the details of the calculation and the explicit forms of the in-plane contributions and out-of-plane hoppings.
In order to model a slab of the material, we cut the block matrix to a finite size, which in this paper is fixed to $51$ layers.
\\On top of this matrix, we introduce a potential $\varphi$, which includes both a contribution from the interfacial charge and a screening contribution from the electrons themselves which populate the interface. Therefore, this component has to be determined self-consistently. In order to do this, we fix the positive charge denisty $\rho$ at the beginning of the slab, and solve the classical equations of the electrodynamics along the $(111)=\hat{Z}$ direction
\begin{equation}
    \partial_Z(\varepsilon(F)\;\partial_Z \varphi)=-\frac{\rho}{\varepsilon_0}
\end{equation}
which for a discrete system of infinite charged planes becomes
\begin{equation}
    \begin{cases}
    \varphi_l=-\frac{a_0}{\sqrt{3}}\sum_{l^\prime=1}^l F_{l^\prime}\\
    \varepsilon_0 \varepsilon(F_l) F_{l} = D_{l}\\
    D_{l} = |e|(n_{2D}-n_{l})
    \end{cases}
    \label{eq:classical_equation}
\end{equation}
where $D_{l}$ is the electric displacement, $F_{l}$ the electric field, $\varepsilon_0$ is the absolute dielectric constant value, $\varepsilon(F)$ is the relative dielectric constant, $n_{2D}$ is the total positive density charge at the interface divided by the in-plane elementary unit cell surface $a_0^2$, while $n_l$ is the 2D density charge on the layer $l$.
Eqs.~(\ref{eq:classical_equation}) involve $\varepsilon$, which leads to solutions which are sensitive to the choice of dielectric constant model. We choose $\varepsilon$ at zero temperature as indicated in Ref.~\cite{bruneel2020electronic} 
\begin{equation}
    \varepsilon(F)=1+\frac{\chi_0}{(1+(\frac{F}{F_0})^2)^{1/3}},
    \label{eq:permittivity}
\end{equation}
where $\chi_0=21000$ and $F_0=80000$ V/m, which for $F=0$ tends to the standard order of magnitude in STO at low temperatures~\cite{neville1972permittivity}.The choice of a $F^{-2/3}$ dependence represents the STO ferroelectric behaviour at low temperatures~\cite{dunitz1957electronic}. This particular behaviour is motivated by the Barret formula~\cite{barret52}. Other parametrization in literature are adopted in Ref.~\cite{neville1972permittivity}.\\
We adopt the following procedure to reach self-consistency: we fix a value of $n_{2D}$ and a trial potential $\varphi_0$, include it in the Hamiltonian and diagonalize it. We find the chemical potential at which the total electron density is $n_{2D}$ and compute the electron density on each layer. We use this density to solve the system~(\ref{eq:classical_equation}) and obtain the potential $\Tilde{\varphi}_0$. At this point the input potential in the Hamiltonian is $\varphi_1 = \alpha \Tilde{\varphi}_0 + (1-\alpha) \varphi_0$, where $\alpha$ is chosen to guarantee a stable convergence. We repeat the procedure until $\varphi_{i+1}\approx\varphi_i$. Typical values of $\alpha$ are $0.8$, $0.9$ and $0.95$, and the stopping criterion is $|\varphi_{i+1}-\varphi_i|<10^{-2}$ eV.
In the following we choose three benchmark choices of $n_{2D}$ ($n_{2D}=1\times10^{14}$~cm$^{-2}$, $n_{2D}=2\times10^{14}$~cm$^{-2}$, and $n_{2D}=3\times10^{14}$~cm$^{-2}$) in order to study changes of the electronic confinement induced by increasing values of electron density. This analysis is rather relevant because the self-consistent densities of 2DEG are in agreement with the expectations, of the order of $1.5\times 10^{14}$ cm$^{-2}$~\cite{walker2014control}).

\subsection{Results in the absence of local Hubbard interaction terms}
\begin{figure}
    \centering
    \includegraphics[width=0.50\textwidth]{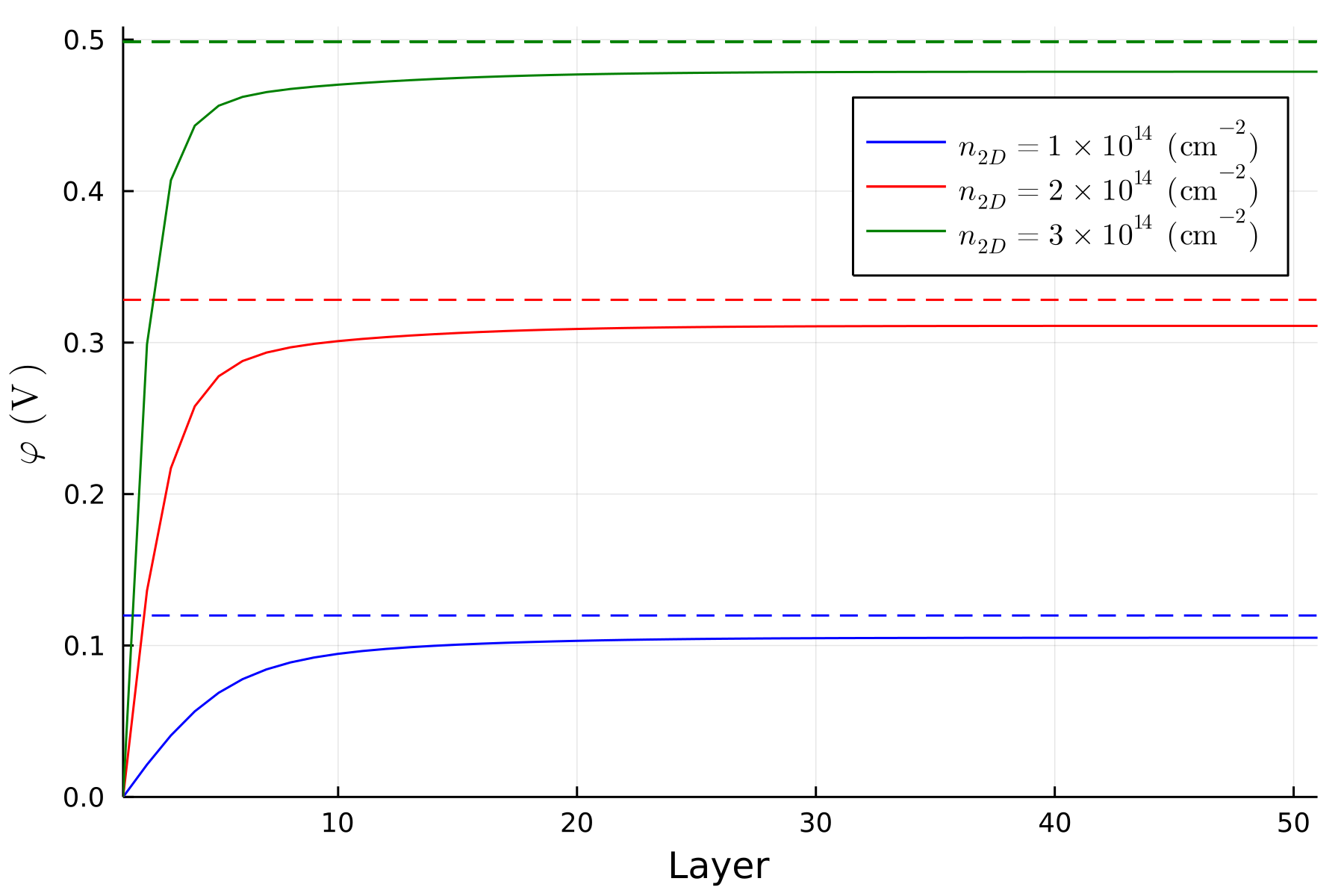}\hfill
    \includegraphics[width=0.50\textwidth]{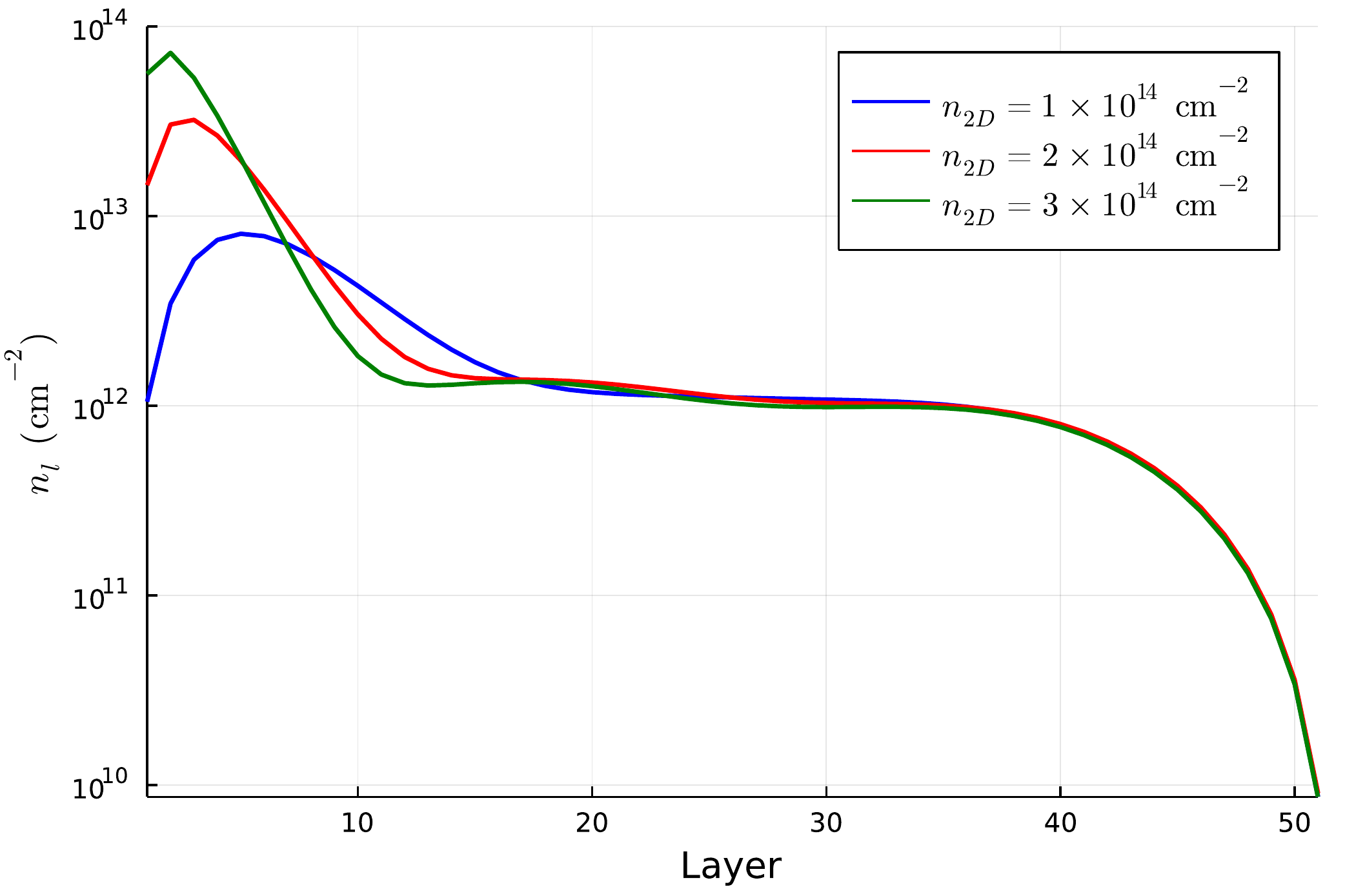}
    \caption{(Left panel) Behaviour of $\varphi$ as a function of the layer position for each benchmark choice of $n_{2D}$. The dashed line represents the corresponding Fermi level. (Right panel) Electron density as a function of the layer position for every benchmark value of $n_{2D}$.}
    \label{fig:potential_and_dens_U0}
\end{figure}

\begin{figure}
    \centering
    \includegraphics[width=0.80\textwidth]{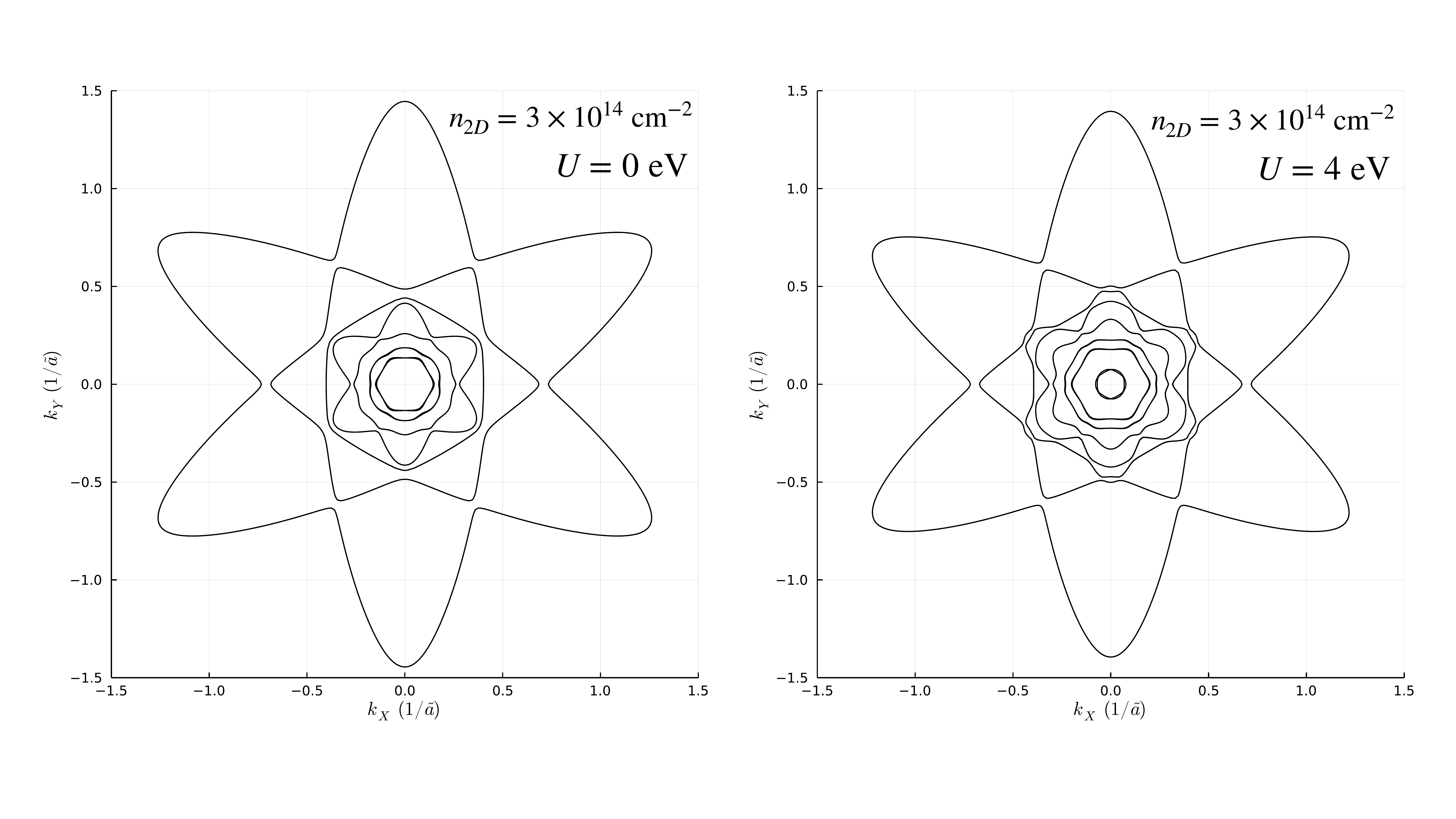}
    \caption{Fermi surfaces for the benchmark choice of $n_{2D}=3\times10^{14}$ cm$^{-2}$ in absence (left panel) and in presence (right panel) of correlations.}
    \label{fig:Fermi_contour}
\end{figure}

In order to clarify the effect of screening, we self-consistently obtain the eigenstates of the Hamiltonian~(\ref{eq:Ham_form}) with the potential of Eqs.~(\ref{eq:classical_equation}) for the three different benchmark parameters defined above. The band structures for low fillings, expressed in terms of the dimensionless quasi-momentum $\vec{k}=\vec{K} a_0 \sqrt{\frac{2}{3}}$, are shown in Fig.~\ref{fig:bands_U0}.
The original $6N$ degrees of freedom are structured in $N$ subsets of $6$ bands each, which are progressively less confined at the interface. By increasing the positive charge at the interface, the splitting between the subsets of bands increases. For $n_{2D}=1\times10^{14}$~cm$^{-2}$, two subsets of bands intersect with one another, while for $n_{2D}=2\times10^{14}$~cm$^{-2}$, and $n_{2D}=3\times10^{14}$~cm$^{-2}$, the first subset of bands is separated from the higher bands. The identification of the subset to which a band belongs is more easily performed by looking at the number of nodes of the wavefunction for each band evaluated at $\vec{k}=0$. We show this result in Appendix~\ref{AppendiceB}. The splitting, and the confinement in turn, is proportional to the slope of the potential close to the first layer. Fig.~\ref{fig:potential_and_dens_U0} shows the potential for all benchmark choices and the 2D charge density for each layer.
Independently of the filling density $n_{2D}$, the chemical potential lies always above the maximum value of the potential well. Therefore, a bulk contribution is always present, as also visible in the right panel of Fig.~\ref{fig:potential_and_dens_U0}. However, the higher is the 2D density, the smaller is the confinement region of the quasi-2DEG, which shrinks to almost 10 layers for $n_{2D}=3\times10^{14}$~cm$^{-2}$. In this case, the 2D charge density of the 2DEG reaches the value of $\sum_{l=1}^{10} n_{l}\approx2.5\times10^{14}$~cm$^{-2}$, the typical order of magnitude for 2DEG densities~\cite{walker2014control}. We point out that the way we extract the density of the 2DEG is different from the one which is used from the typical (001) interface, since in the (111) interface, the first band is not really 2D due to the fact that the electron hopping happens between the different planes. Therefore, the estimation of the 2DEG density computed as the area of the FS of the external band is naive.
An interesting feature which appears above the chemical potential is the band splitting at $\vec{k}=0$ for the higher bands. This splitting disappears by removing the atomic SOC and thus it resembles a huge linear Rashba splitting induced by the combined effect of the electric potential, which naturally breaks the inversion symmetry, and the atomic SOC. Even if the Fermi energy is at filling energy lower than these splitting, the chemical potential can be changed by using an external gate voltage, in order to investigate these kind of bands. The application of such a gate does not change the splitting between the bands, since this is mostly influenced by the positive charge at the interface.\\ In Fig.~\ref{fig:Fermi_contour} we show the FSs of the band for the benchmark density $n_{2D}=3\times10^{14}$ cm$^{-2}$. The contour shows a six-fold symmetry of the energy spectra and, in agreement with the analysis of Ref.~\cite{walker2014control,trama2022gate,trama2022tunable}, every ellipse-shaped band possesses a strong $d$-orbital character away from $\vec{k}\sim0$, while nearby the $\Gamma$ point the $a_{g}$ or $e_{g}^\pi$ character of the band is restored (i.e. the spherical symmetry of the inner bands), due to the trigonal crystal field.

\subsection{Effect of local Hubbard interaction terms}
\begin{figure}
    \centering
    \includegraphics[width=0.62\textwidth]{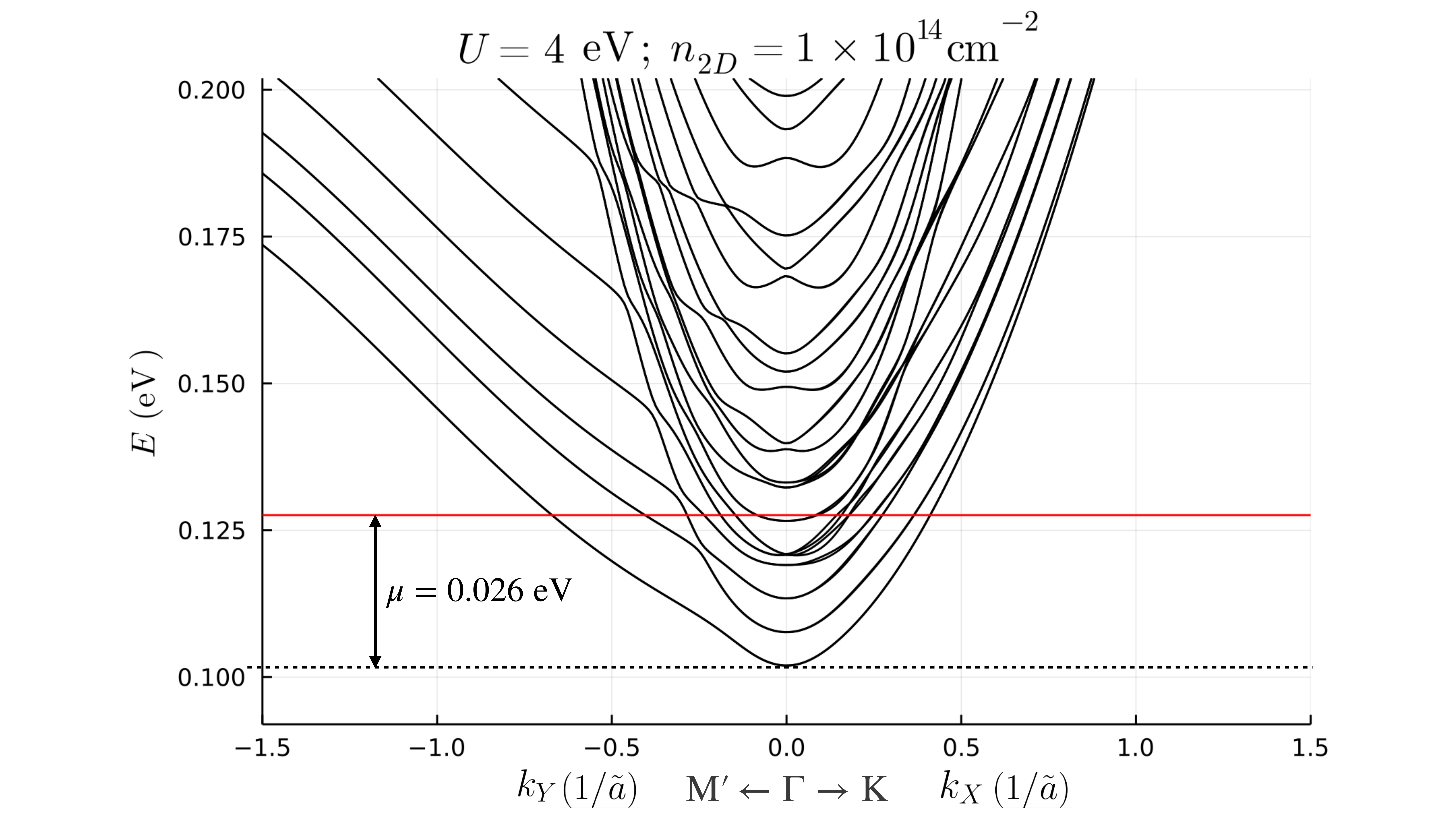}\\
    \includegraphics[width=0.62\textwidth]{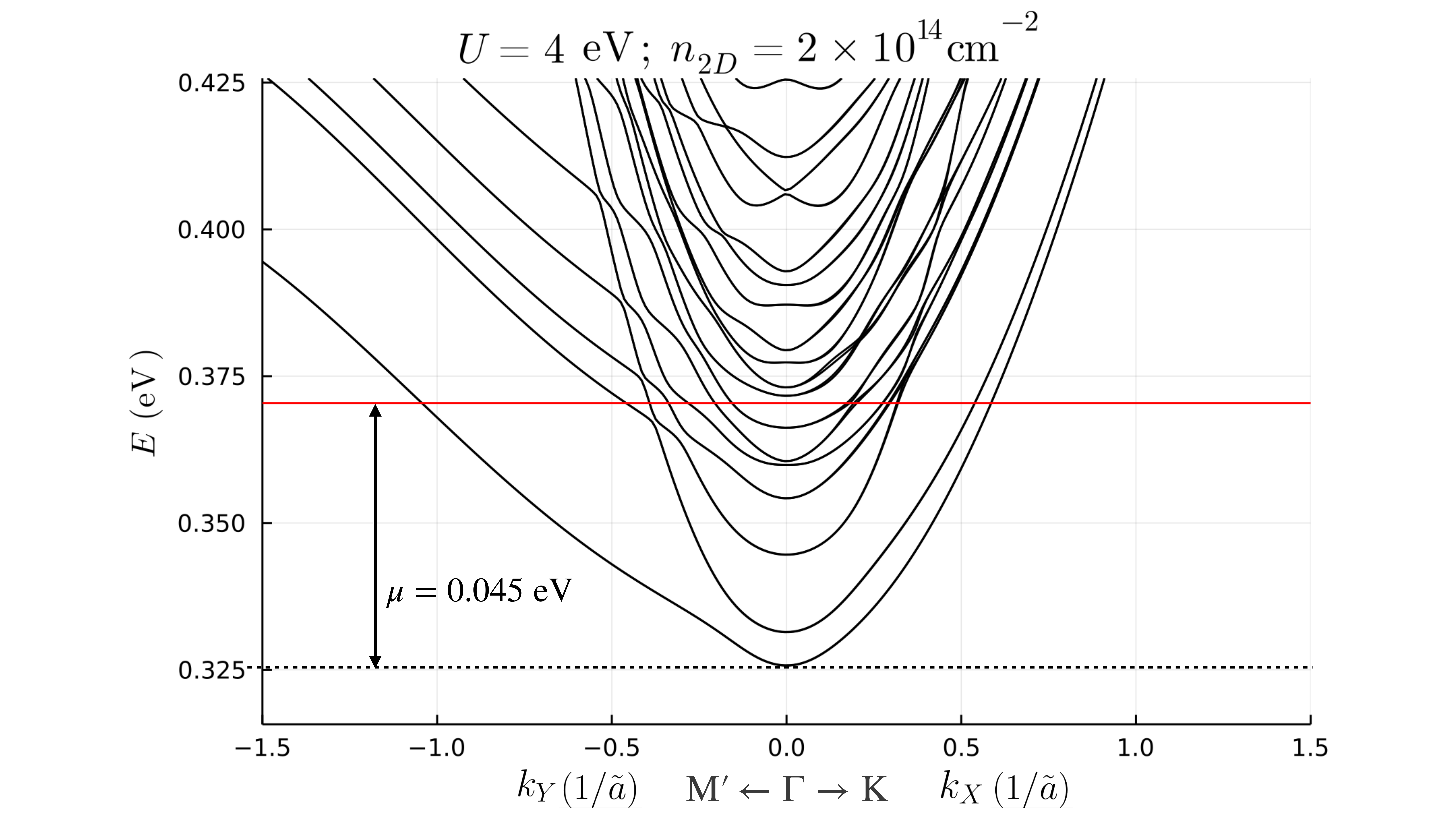}\\\
    \includegraphics[width=0.62\textwidth]{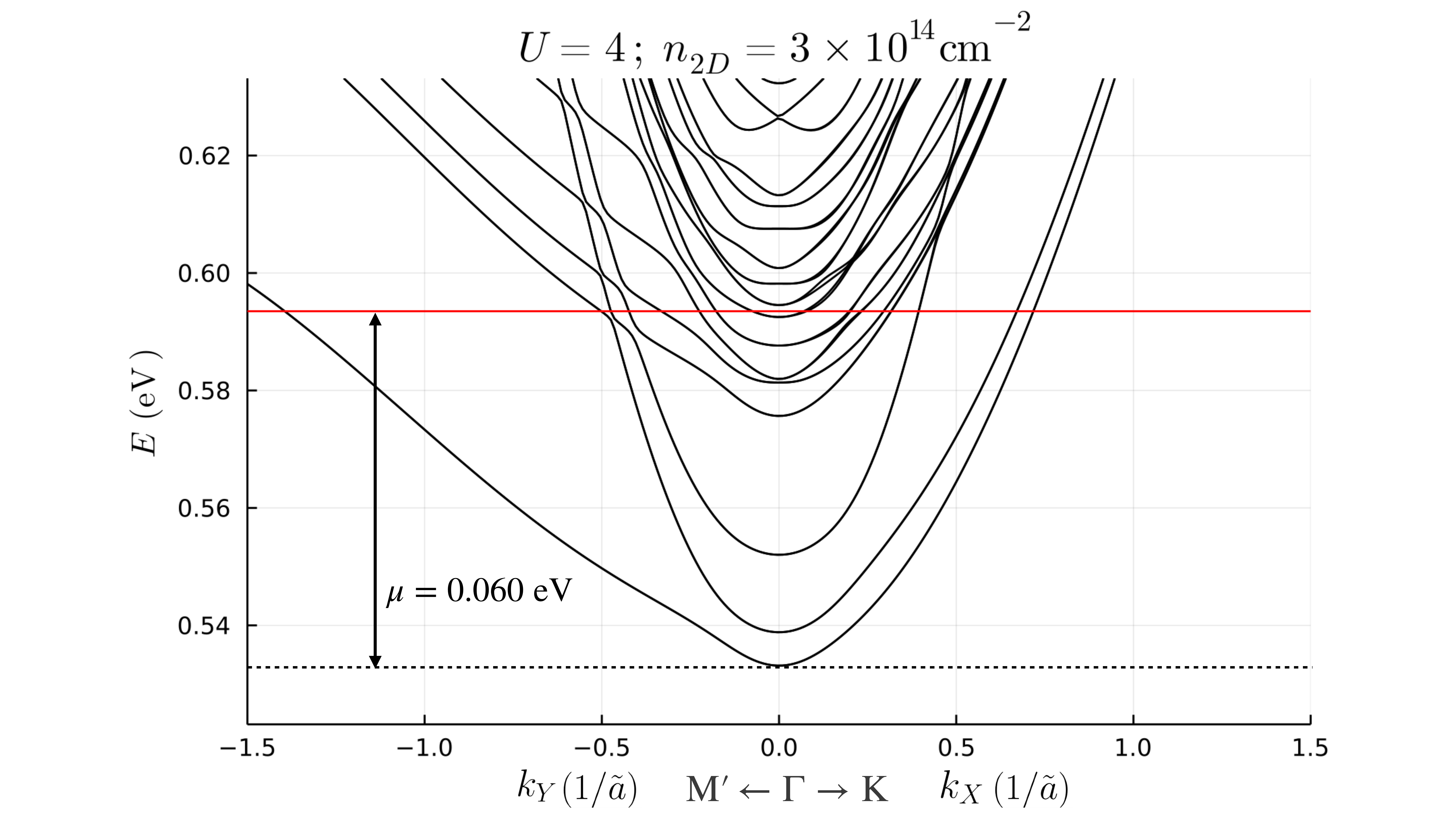}
    \caption{Low filling electronic band structure in presence of Coulomb interactions for the benchmark choice of density $n_{2D}=1\times10^{14}$ cm$^{-2}$ (upper panel), $n_{2D}=1\times10^{14}$ cm$^{-2}$ (middle panel) and $n_{2D}=3\times10^{14}$ cm$^{-2}$ (lower panel). The red line is the Fermi level of the system.}
    \label{fig:bands_U4}
\end{figure}
\begin{figure}[b]
    \centering    
    \includegraphics[width=0.65\textwidth]{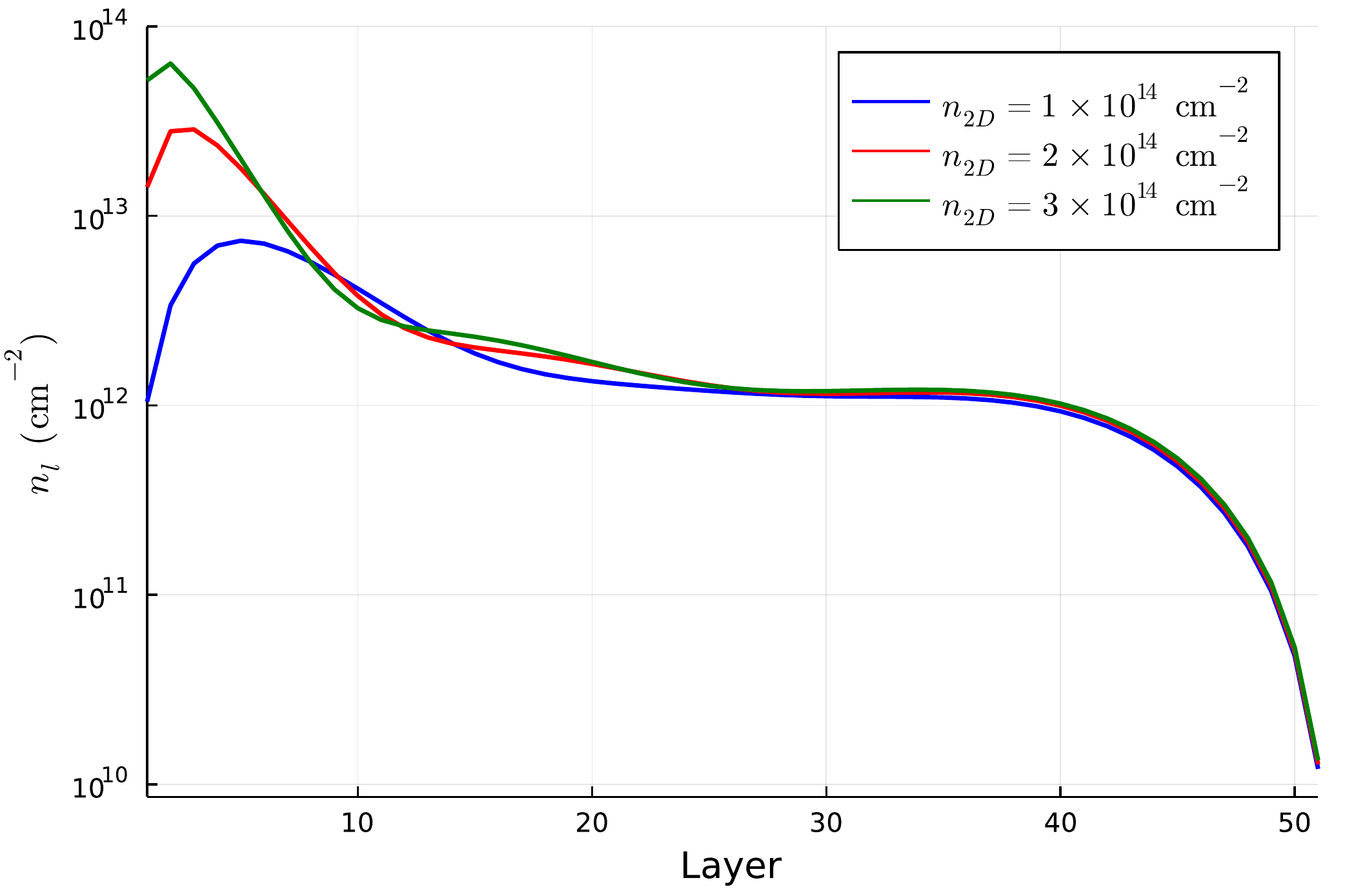}
    \caption{Electron density as a function of the layer position for every benchmark value of $n_{2D}$ in presence of Coulomb interactions.}
    \label{fig:U4_dens}
\end{figure}

In this section we explore the effect of local Hubbard electron-electron terms on the result we presented above. These interactions can be expressed in a mean-field approximation as
\begin{equation}
    H_{\rm{C}}=\sum_{\vec{K}}\sum_{\alpha,l}n_{\alpha\uparrow,l,\vec{K}}\left(U\nmean{\alpha\downarrow,l}+\sum_{\beta\neq\alpha}U^{\prime}\nmean{\beta\uparrow,l}+U^{\prime}\nmean{\beta\downarrow,l}\right)+(\uparrow\leftrightarrow\downarrow),
    \label{eq:Coulomb}
\end{equation}
where $n_{\alpha\uparrow,l,\vec{K}}=d_{\alpha\uparrow,l,\vec{K}}^\dagger d_{\alpha\uparrow,l,\vec{K}}$, we exploited spatial homogeneity in the interfacial plane so that mean densities are independent of positions, $\alpha$ runs over the orbital degree of freedom, and $U$ and $U^\prime$ parametrize the strengths of the interaction. In the absence of any term breaking the $C_{3v}$ symmetry and the time-reversal invariance, for each layer the electron density is equal for every orbital and spin.
In such a regime, the $U^\prime$ terms provides only a renormalization of the 
$U$ term; therefore, we can neglect them reducing at the same time the computational effort for the simultaneous self-consistent calculation of the local Hubbard potential $U \langle n_l\rangle$ and the potential $\varphi_l$. Eq.~(\ref{eq:Coulomb}) shows that the the local Coulomb interaction introduces an effective potential varying over each layer, proportional to the local particle density at that layer. We expect that this leads to a broadening of the electron density over the whole slab of material, since it favours energetically the lowest occupied layers. \\ 
We choose a benchmark value of $U=4$~eV, as chosen in Ref.~\cite{monteiro2019band}, and compute the bands shown in Fig.~\ref{fig:bands_U4} for the same benchmark values of $n_{2D}$, in order to clearly discriminate the effects originating from local Hubbard interactions on the band structure.
For $n_{2D}=1\times10^{14}$~cm$^{-2}$ and $n_{2D}=2\times10^{14}$~cm$^{-2}$, the band structures change only slightly from the situations without Hubbard terms. However, at $n_{2D}=3\times10^{14}$~cm$^{-2}$ the local Hubbard terms significantly enhance the separation between the confined and the free bands. Moreover, due to the effect of the Hubbard terms, an additional sub-band crosses the chemical potential close to the $\Gamma$ point. As a result, as shown in Fig.~\ref{fig:Fermi_contour}, the FS shows a reconfiguration in the same region of the Brillouin zone.    
\begin{figure}
    \centering
    \includegraphics[width=0.55\textwidth]{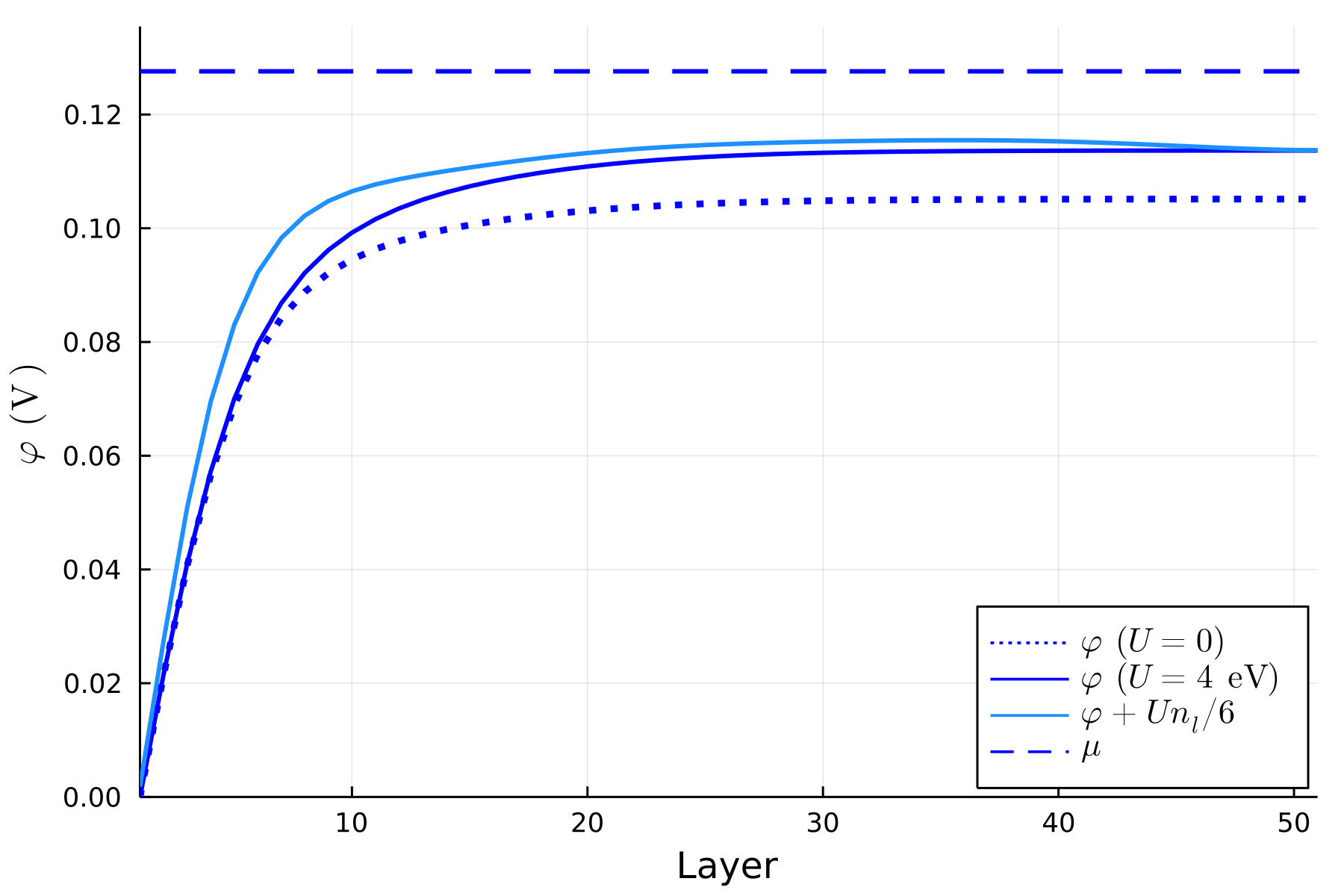}\\
    \includegraphics[width=0.55\textwidth]{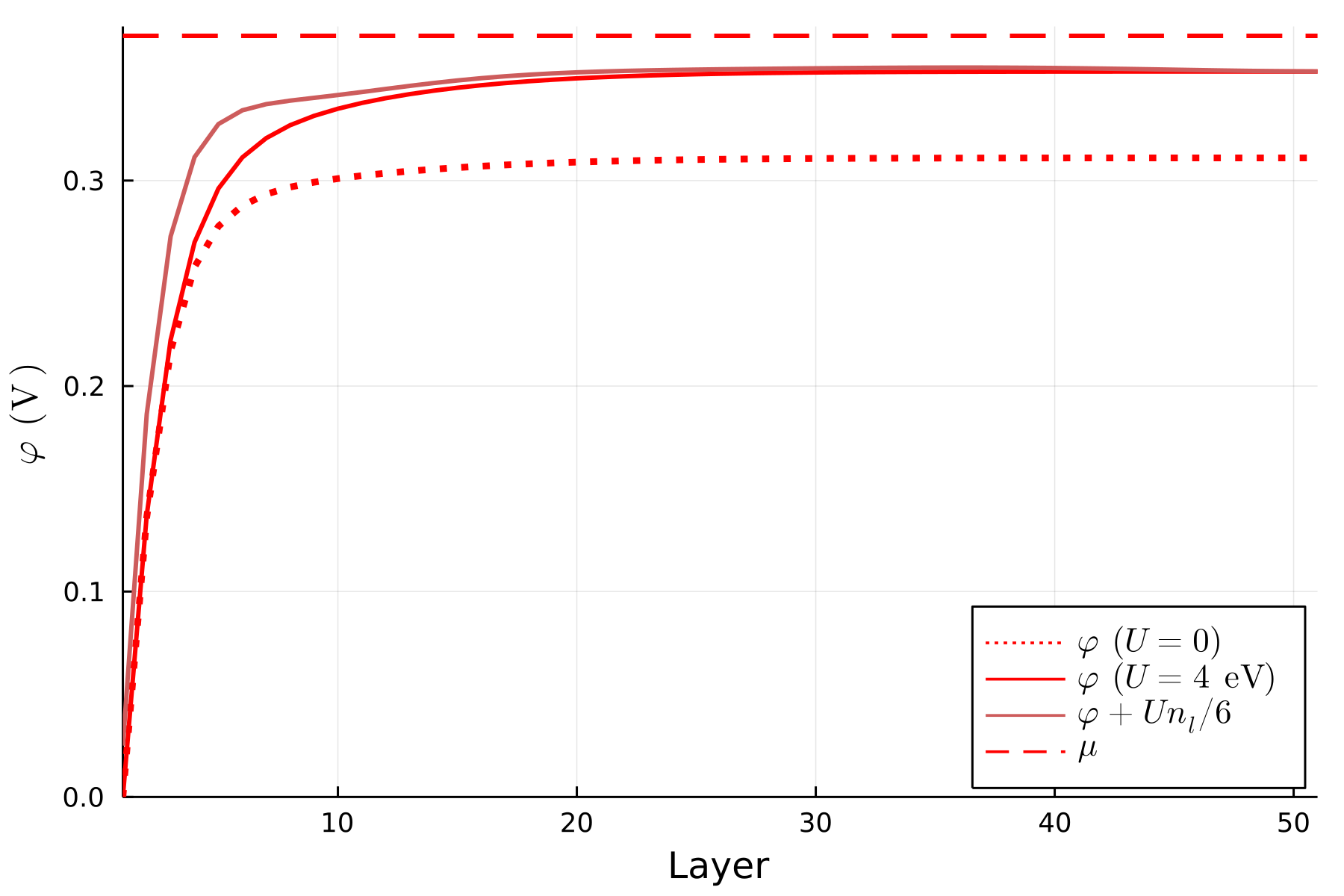}\\
    \includegraphics[width=0.55\textwidth]{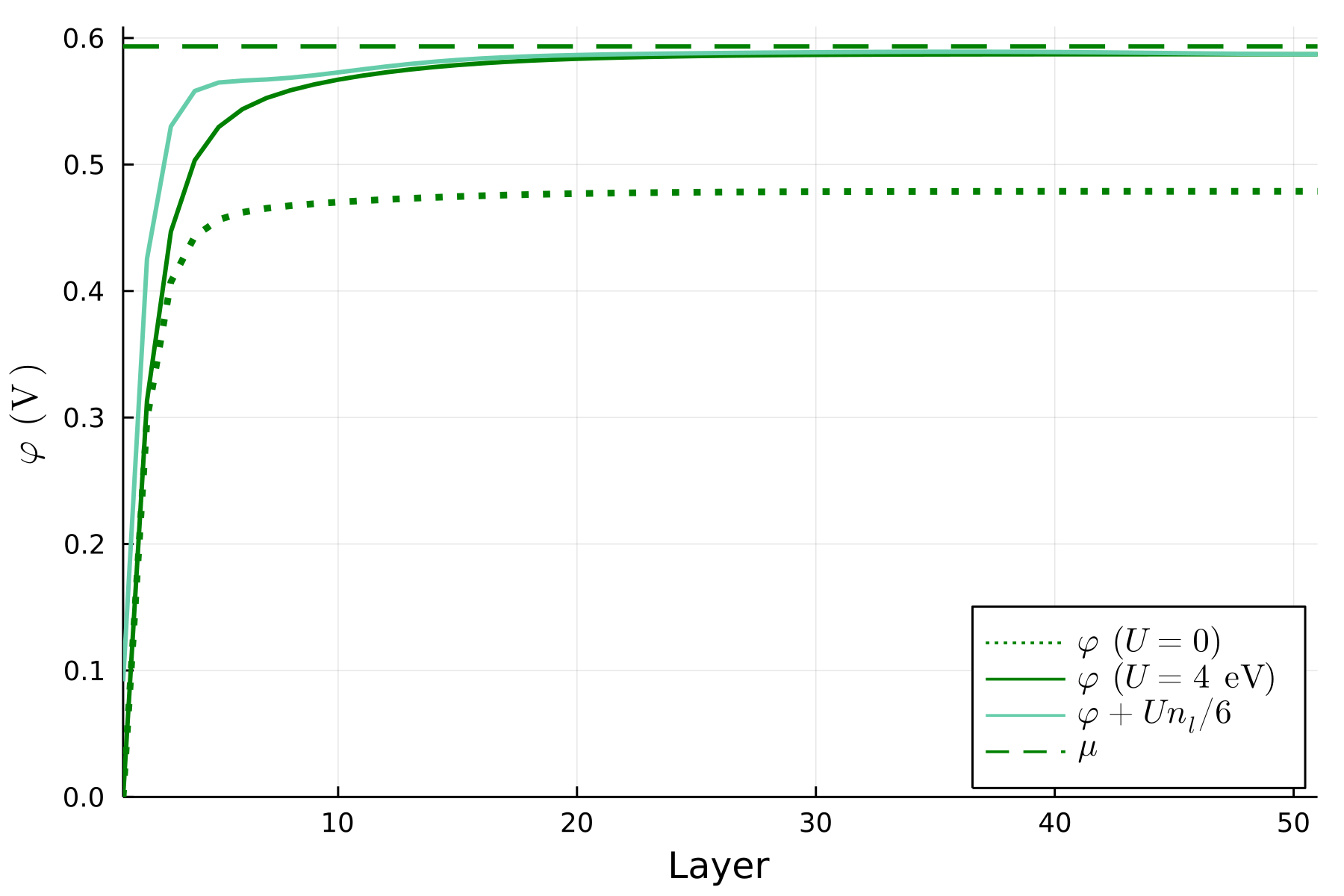}
    \caption{Behaviour of $\varphi$ and $\varphi + U n_{l}/6$ as a function of the layer position in the presence of Coulomb interactions for $n_{2D}=1\times10^{14}$ cm$^{-2}$ (upper panel), $n_{2D}=2\times10^{14}$ cm$^{-2}$ (middle panel) and $n_{2D}=3\times10^{14}$ cm$^{-2}$ (lower panel). The dashed line represents the corresponding Fermi level, while the dotted line is the corresponding potential in absence of Coulomb interactions.}
    \label{fig:poten_U4}
\end{figure}
We can clarify interesting features of the electronic band structure by comparing the self-consistent potential with and without the effects of Hubbard terms, in the former case accounting also for the effective potential $U \langle n_l\rangle=U n_l/6$. We show these results in Fig.~\ref{fig:poten_U4}. For $n_{2D}=3\times10^{14}$~cm$^{-2}$, the effective potential exhibits a small peak which enhances the separation between the high-slope region to the plateaux.
Even in this case, however, the chemical potential is above the highest value of the potential, which results in a bulk component. The electron density for the case of $U=4$ eV is reported in Fig.~\ref{fig:U4_dens}. Actually, it shows an intermediate region between the layers $10$ and $30$ where the local Hubbard terms induce an enhancement of the local density. This is a very interesting result since the Hubbard terms does not deplete the 2DEG but adds a modulation of the electron density in the intermediate region. This is mainly due to the fact that the interfacial electron charge concentration repels the electrons in the intermediate region, creating a slight inflexion in the effective potential felt by each electron and mildly favoring occupation of the intermediate region.

\section{Discussion and conclusions}
\label{sec:discussion}
In this manuscript we have systematically discussed a TB supercell method to describe the band structure of (111) LAO/STO interface taking into account the effect of atomic SOC, trigonal strain, the electronic confinement and of local Hubbard electron-electron interactions within a fully self-consistent procedure. The sub-band energy and the FS show full agreement with the observed electronic structure by ARPES~\cite{walker2014control} and describes how the 2DEG arises from the quantum confinement of $t_{2g}$ electrons near the surface due to band bending potential, for different values of the positive charge density at the interface, i.e. by varying the concentration of the oxygen vacancies.
In particular, we have shown how the effect of local Hubbard electron-electron interactions changes the density distribution over the layers close to the surface. Indeed, the 2DEG at interface is not depleted by local Hubbard  interactions which, instead, induce a modulation of the electron density as a function of the layers number. The net effect of the Hubbard interactions is to enhance the electron density in the intermediate spatial region between the first layers at the interface and the bulk. 
This effect was previously analyzed in Ref.~\cite{delugas2011spontaneous} for the (001) interface for which the Coulomb interactions have the effect of helping the formation of the 2DEG. The main difference compared to the (001) is the orbital structure of the electrons at the interface which causes a different impact of the Coulomb interactions on the results. 
However, we do not find any qualitative changes in the behaviour of the eigenstates in the confined region due to the effect of the Hubbard interaction as found in Ref.~\cite{monteiro2019band}. This is possibly due to different numerical strategies used in the band structure computation. Moreover, despite in our model a Rashba coupling or any odd hopping parameters in the quasi-momentum was not inserted explicitly, an interesting huge linear splitting for the higher bands above the Fermi level appears, probably due to the combined effect of SOC and the electric potential. Its origin can be the subject of a further study.

\section*{acknowledgments}
C.A.P. acknowledges support from Italy’s MIUR PRIN project TOP-SPIN (Grant No. PRIN 20177SL7HC).
\clearpage
\appendix
\section{Model}\label{AppendiceA}
In this appendix we summarize the procedure for obtaining the TB supercell Hamiltonian and than we provide an analytical estimation of an approximate solution for the electronic confinement using the Poisson-Schr\"{o}dinger approximation.
\subsection{Tight-binding supercell Hamiltonian}
The bulk Hamiltonian in the (001) coordinates is described in Eq.~(\ref{eq:bulk_ham}). A rotation of the system of coordinates in the (111) direction leads to the following transformation
\begin{equation}
    \begin{cases}
    \hat{x}=\frac{1}{\sqrt{6}}\left(-\sqrt{3}\hat{e}_{\bar{1}10}-\hat{e}_{\bar{1}\bar{1}2}+\sqrt{2} \hat{e}_{111}\right),\\
    \hat{y}=\frac{1}{\sqrt{6}}\left(\sqrt{3}\hat{e}_{\bar{1}10}-\hat{e}_{\bar{1}\bar{1}2}+\sqrt{2} \hat{e}_{111}\right),\\
    \hat{z}=\frac{1}{\sqrt{3}}\left(\sqrt{2}\hat{e}_{\bar{1}\bar{1}2}+ \hat{e}_{111}\right).
    \end{cases}
    \label{eq:sys_coo}
\end{equation}
Once the corresponding $K_i$ is substituted in Eq.~(\ref{eq:bulk_ham}), one can identify
\begin{equation}
    \begin{cases}
    A^\dagger=e^{i\frac{K_{111}\Tilde{a}}{\sqrt{2}}},\\
    A=e^{-i\frac{K_{111}\Tilde{a}}{\sqrt{2}}},
    \end{cases}
\end{equation}
where $\Tilde{a}=\sqrt{\frac{2}{3}}a_0$ is the lattice parameter projected in the (111) plane, $A^\dagger$ is the jump operator along the $\hat{e}_{111}$ direction and $A$ its conjugate.
Therefore we obtain
\begin{equation}
    H_{\textrm{TB}}=H_t A^{\dagger} + H_t^{\dagger} A,
    \label{eq:H_tb_dagger}
\end{equation}
where
\begin{equation}
    H_t=
    \begin{pmatrix}
        \epsilon_{yz} & 0 & 0 \\
         0 & \epsilon_{zx} & 0 \\
         0 & 0 & \epsilon_{xy} 
    \end{pmatrix},
\end{equation}
having neglected the spin degree of freedom, with
\begin{linenomath}
\begin{eqnarray}
     \nonumber &\epsilon_{yz}&=-t_D\left(e^{i k_Y}+e^{i(\frac{\sqrt{3}}{2}k_X-\frac{1}{2}k_Y)}\right)-t_Ie^{-i(\frac{\sqrt{3}}{2}k_X+\frac{1}{2}k_Y)},\\
     &\epsilon_{zx}&=-t_D\left(e^{i k_Y}+e^{-i(\frac{\sqrt{3}}{2}k_X+\frac{1}{2}k_Y)}\right)-t_Ie^{i(\frac{\sqrt{3}}{2}k_X-\frac{1}{2}k_Y)},\\
     \nonumber &\epsilon_{xy}&=-t_D\left(e^{-i(\frac{\sqrt{3}}{2}k_X+\frac{1}{2}k_Y)}+e^{i(\frac{\sqrt{3}}{2}k_X-\frac{1}{2}k_Y)}\right)-t_I e^{i k_Y}.
     \label{interlayer}
\end{eqnarray}
\end{linenomath}

Here we have introduced the dimensionless quasi-momentum $\vec{k}=\vec{K}\Tilde{a}$ and called $\hat{X}=(\Bar{1}10)$ and $\hat{Y}=(\Bar{1}\Bar{1}2)$. The direct $t_D$ and indirect $t_I$ couplings have been fixed to the values $t_D=0.25$ eV and $t_I=0.02$ eV~\cite{trama2021straininduced} via comparison with angular resolved photoemission spectroscopy data. 
The Tight-Binding supercell matrix is obtained by imposing some boundary conditions at the extrema of the lattice of layers. We include open boundary conditions for a slab of 51 layers. In this way, we obtain the finite form of Eq.~(\ref{eq:Ham_form}) of the main text.
\\As discussed in the main text, the local terms are contained in $H_0=H_{\rm{SO}}+ H_{\text{TRI}}$. 
\\$H_{\rm{SO}}$ is the atomic SOC coupling, which has the following expression
\begin{linenomath}
\begin{equation}
    H_{\text{SOC}}=\frac{\lambda}{2}\sum_{\vek}\sum_{ijk,\sigma\sigma'}i\varepsilon_{ijk}
    d_{i\sigma,\vek}^{\dagger} \sigma^{k}_{\sigma\sigma'}d_{j\sigma',\vek}
    \label{eq:spinorbit}
\end{equation}
\end{linenomath}
where $\varepsilon_{ijk}$ is the Levi-Civita tensor and $\{i,j,k\}$ runs over the label $\{yz,zx,xy\}$, and $\sigma^k$ are the Pauli matrices. We fix the SOC coupling $\lambda=0.01$ eV, as a typical order of magnitude~\cite{monteiro2019band}.
\\The trigonal crystal field Hamiltonian $H_{\rm{TRI}}$ takes into account the strain at the interface along the (111) direction. The physical origin of this strain is the possible contraction or dilatation of the crystalline planes along the (111) direction. This coupling has the form~\cite{khomskii2014transition}
\begin{linenomath}
\begin{equation}
    H_{\text{TRI}}=\frac{\Delta}{2}\sum_{\vek}\sum_{i\neq j,\sigma} d_{i\sigma,\vek}^{\dagger} d_{j\sigma,\vek}.
    \label{eq:trigonal}
\end{equation}
\end{linenomath}
We fix $\Delta=-0.005$ eV as reported in~\cite{de2018symmetry}.
\\The last term included in our model is the local electrostatic potential $\varphi_l$. This has the form of a diagonal matrix which has the same entry for each layer. Here we report the form of the full Hamiltonian in a block matrix form
\begin{equation}
    H=\begin{pmatrix}
        H_0+\varphi_1 & H_t & 0 & 0 & 0 &...\\
        H_t^{\dagger} & H_0+\varphi_2 & H_t & 0 & 0 & ...\\
        0 & H_t^{\dagger} & H_0+\varphi_3 & H_t & 0 & ...\\
        0 & 0 & H_t^{\dagger} & H_0+\varphi_4 & H_t & ...\\
        ... & ... & ... & ... & ... & ...\\
    \end{pmatrix}.
    \label{eq:Ham_form_complete}
\end{equation}

\subsection{Poisson-Schr\"{o}dinger approximation}
\begin{figure}
    \centering
    \includegraphics{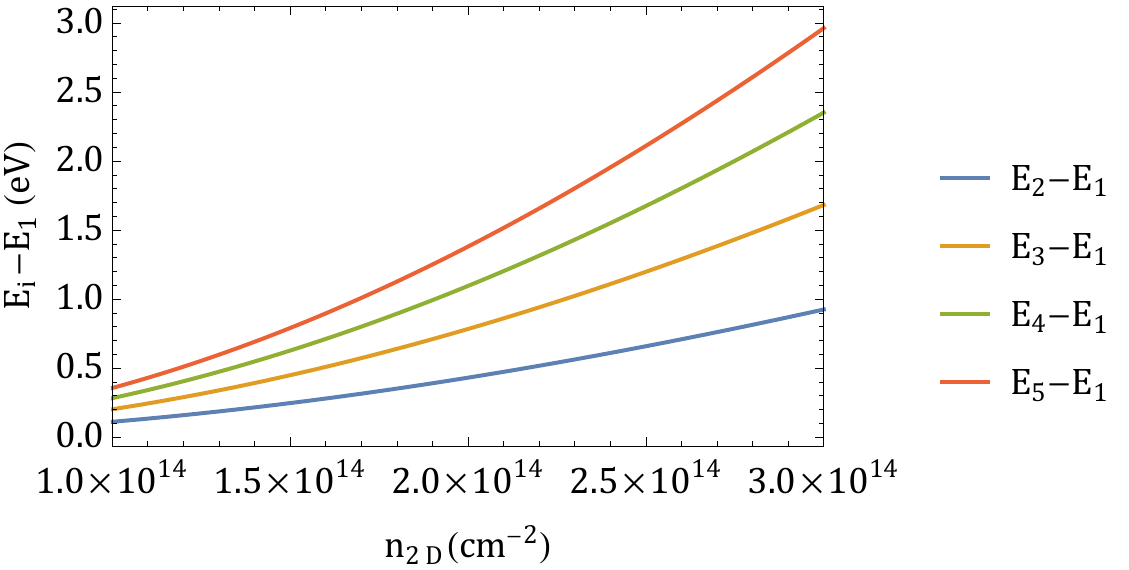}
    \caption{Difference between the first and the next out-of-plane eigenvalues evaluated through the analytical approach as a function of the positive density charge at the interface.}
    \label{fig:splitting_dif}
\end{figure}

The Poisson-Schr\"{o}dinger approximation consists in performing an expansion to lowest order of the jump operators as
\begin{equation}
\begin{cases}
     A=1-i\kappa-\frac{\kappa^2}{2},\\
     A^\dagger=1+i\kappa-\frac{\kappa^2}{2},
\end{cases}
\end{equation}
where for simplicity we defined $\kappa=K_{111}\Tilde{a}/\sqrt{2}$, and we than make the correspondence of $k\rightarrow -i\frac{\partial}{\partial l}$, where $l$ is a continuous coordinate which for integer numbers indicates the number of layer.
The Hamiltonian~(\ref{eq:H_tb_dagger}) becomes the following
\begin{equation}
    H_{\textrm{TB}}= 2\text{Re}{(H_t)}\left(1-\frac{\kappa^2}{2}\right)-2\text{Im}({H_t}) \kappa.
\end{equation}
We will now consider as independent the in-plane dispersion consider from the out-of-plane part. For in-plane quasi-momenta sufficiently close to zero, we write 
\begin{equation}
    H_{\text{TB}}= 2\text{Re}{(H_t)}+\left(\frac{\text{Im}{(H_t)}^2}{\text{Re}({H_t})}\right)-2\text{Re}{(H_t(0,0))}q^2,
\end{equation}
where $q=\kappa+\frac{\text{Im}{(H_t)}}{\text{Re}({H_t})}$ and $H_t(0,0)=(2t_D+t_I)$ for all the orbitals.
Here we identify the terms independent from $q$ as the in-plane Hamiltonian, while the $q^2$ term is the out-of-plane dispersion. 
Therefore the out-of-plane part of the Hamiltonian is
\begin{equation}
    H_{Z}=-\text{Re}(H_t(0,0))\left(-i\frac{\partial}{\partial l}+\frac{\text{Im}(H_t(0,0))}{\text{Re}(H_t(0,0))}\right)^2=-(2t_D+t_I)\frac{\partial^2}{\partial^2 l}.
\end{equation}
To this Hamiltonian we add the one-body potential $V(l)$ which is unknown. The potential $V(l)$ is the potential acting on the electrons which are attracted by the positive charge at the interface. Combining the classical equations of the electromagnetism with the Schr\"{o}dinger equation we can solve the out-of-plane part of the problem as
\begin{equation}
    \begin{cases}
    -(2t_D+t_I)\frac{\partial^2\psi}{\partial^2 l}+V(l)\psi=E\psi\\
    F=-\partial_l V\\
    F\varepsilon(F)\varepsilon_0=D\\
    \partial_l D= \rho
    \end{cases}
\end{equation}
where $F$ is the electric field generated by $V$, $D$ is the electric displacement, $\varepsilon_0$ is the vacuum permittivity, while $\varepsilon(F)$ is the relative permittivity, and $\rho$ is the 3D charge density. 
\\In a self-consistent approach we would solve the equations numerically since $V(l)$ is determined by the electron charge density on each layer, which reflects also the screening by $\varepsilon(F)$. By doing this, the potential naturally bends until the electric field goes to zero at the end of the slab of the material.
In this section, however we want to quantify the confinement effects by studying their influence on the lowest energy levels. In this case we will find the solutions of the infinite potential well whose slope is the slope of a realistic potential for $l\to 0$.
\\By fixing a value of $\rho=|e|n_{2D}\sqrt{2}/\Tilde{a}$, we find $D=|e|n_{2D}$. Moreover, in this approximation $V(l)=F l \frac{\Tilde{a}}{\sqrt{2}}$ (expressed in V). Let us solve the Schrödinger equation
\begin{equation}
    -\frac{\partial^2\psi}{\partial^2 l}+\left(\frac{V(l)-E}{-(2t_D+t_I)}\right)\psi=0.
\end{equation}
By choosing
\begin{equation}
    \begin{cases}
        \gamma=\frac{E\sqrt{2}}{\Tilde{a}F},\\
        v=\frac{F\Tilde{a}}{\sqrt{2}},\\
        \xi=\frac{\Tilde{a}F}{\sqrt{2}\;(2t_D+t_I)},
    \end{cases}
\end{equation}
we obtain
\begin{equation}
    -\frac{\partial^2\psi}{\partial^2 l}+\xi\left(l-\gamma\right)\psi=0.
\end{equation}
From this equation we see that the confinement exists until $l=\gamma$.
The equation is now clearly an Airy equation with the following boundary conditions
\begin{equation}
    \begin{cases}
    \psi(l\to\infty)=0\\
    \psi(l=0)=\psi(-\gamma\xi^{1/3})=0.
    \end{cases}
\end{equation}
From the last condition we find the eigenvalues (contained in $\gamma$) which are the zeros of the Airy's function $\mathcal{Z}_i$ as
\begin{equation}
    E_i=-\mathcal{Z}_i v^{2/3} (2t_D+t_I)^{1/3}.
\end{equation}
In order to include the confinement effects in a simple way, let us suppose to take the dielectric permittivity in Eq.~(\ref{eq:permittivity}).
By choosing $n_{2D}=1\times10^{14}$ cm$^{-2}$ we obtain $F=9.84\times10^{7}$ V/m. We find from the relation $l_i=E_i/v$ that we confined the system over $7$ layers. From Fig.~\ref{fig:potential_and_dens_U0} we see that the peak of the density is located approximately between the layer 5 and 7, which is in agreement with the analytical estimation.
The splitting between the first eigenvalue and the subequent (which gives the splitting between the in-plane sub-bands), is depicted in Fig.~\ref{fig:splitting_dif} by varying the positive charge density at the interface. However by comparing the splitting of the self-consistent bands in Fig.~\ref{fig:bands_U0}, we see that the predicted splitting is an order of magnitude higher that the self-consistent one.

\section{Sub-set band behaviour}\label{AppendiceB}
\begin{figure}
    \centering
    \includegraphics[width=0.45\textwidth]{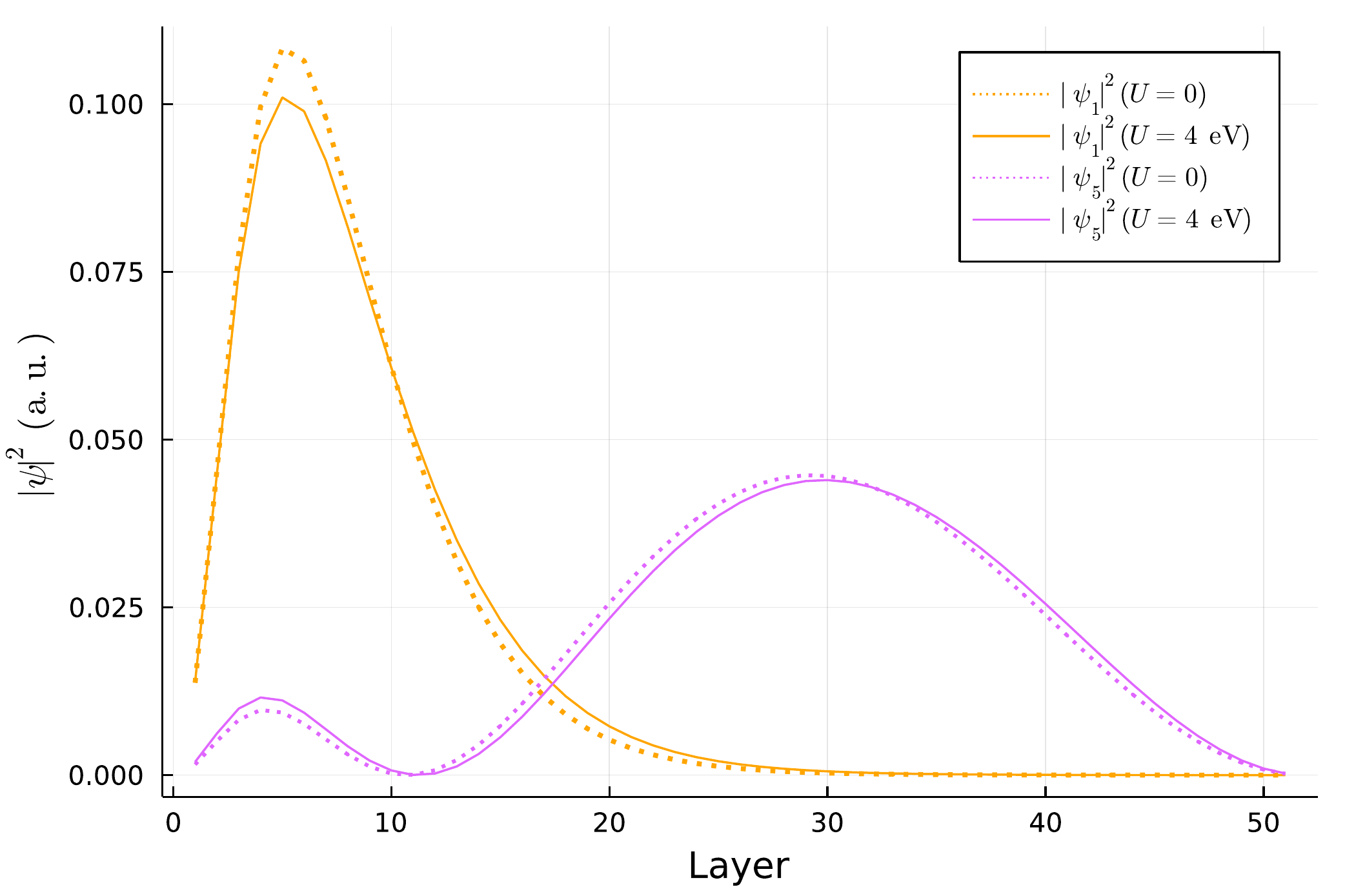}\hfill \includegraphics[width=0.45\textwidth]{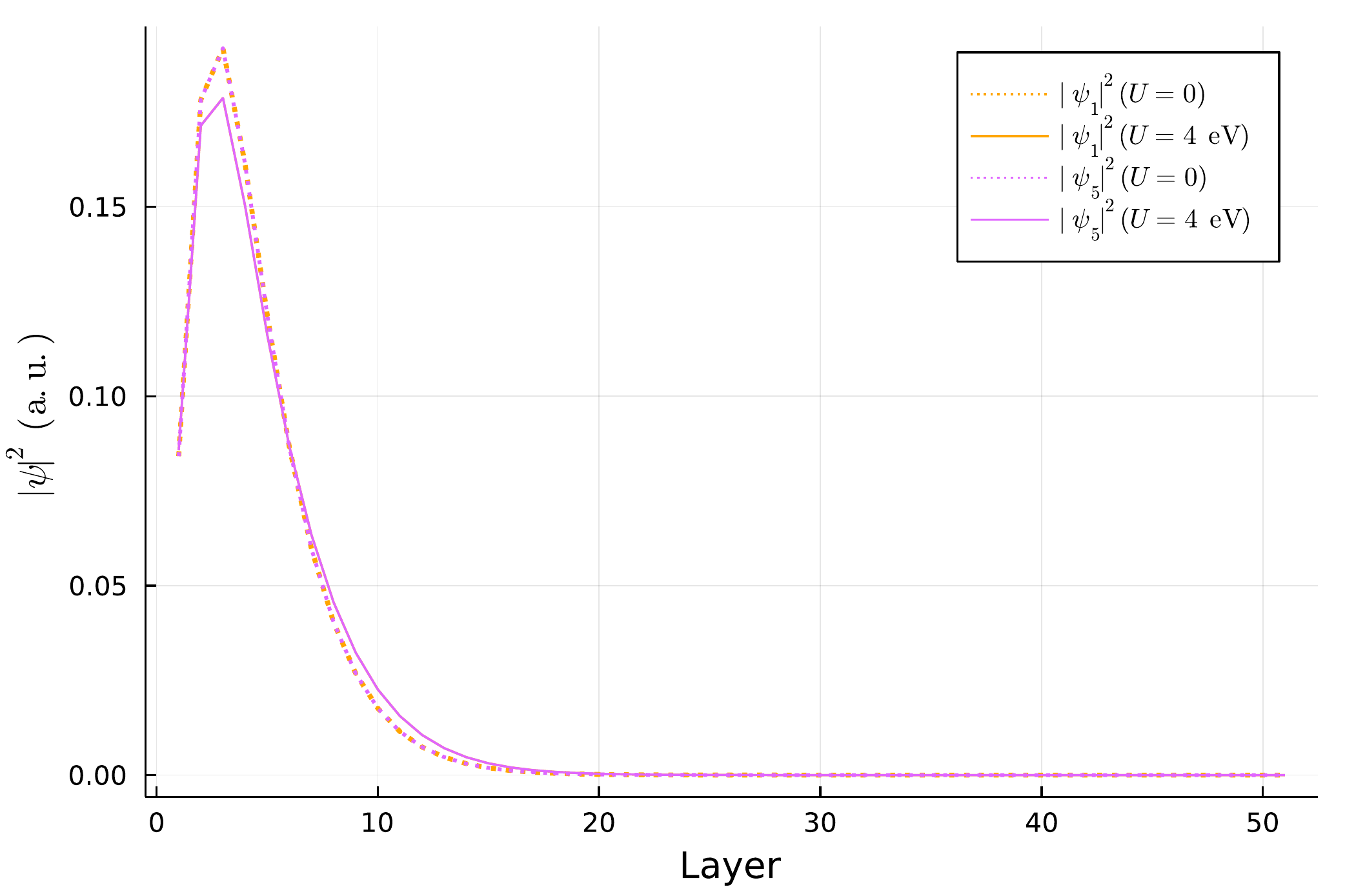}
    \caption{Spatial distribution of the eigenstates evaluated for $\vec{k}=0$, for benchmark choice of $n_{2D}=1\times10^{14}$ cm$^{-2}$ (left panel) and $n_{2D}=2\times10^{14}$ cm$^{-2}$ (right panel).}
    \label{fig:psik0}
\end{figure}

In the case of continuous quantum systems, it is well known that the ordering of the eigenstates can be identified by the number of nodes of the corresponding eigenfunctions. Therefore, in order to classify the sub-band character of the first occupied bands, in Fig.~\ref{fig:psik0} we show the spatial occupation for the eigenstates of the Hamiltonian evaluated for $\vec{k}=0$. In figure, we highlight the first and the fifth bands for the benchmark choice of $n_{2D}=1\times10^{14}$ cm$^{-2}$ and $n_{2D}=2\times10^{14}$ cm$^{-2}$. In the first case, the fifth band has a single node, since it belongs to the second sub-set of bands. This behaviour can be explained in a simple scenario of separation of variables between the in-plane and the out-of-plane wavefunction, for which the splitting induced by the confinement is smaller than the splitting induced by SOC and the trigonal crystal field. The behaviour changes for larger $n_{2D}$, since the slope of the potential increases and in turn also the sub-bands splitting. The case of $n_{2D}=3\times10^{14}$ cm$^{-2}$ is not shown since the qualitative behaviour is the same as $n_{2D}=2\times10^{14}$ cm$^{-2}$. In all the cases we have studied, the correlations do not induce any modification of the spatial distribution of the system eigenstates.



\bibliography{Bib}

\begin{thebibliography}{24}%
\makeatletter
\providecommand \@ifxundefined [1]{%
 \@ifx{#1\undefined}
}%
\providecommand \@ifnum [1]{%
 \ifnum #1\expandafter \@firstoftwo
 \else \expandafter \@secondoftwo
 \fi
}%
\providecommand \@ifx [1]{%
 \ifx #1\expandafter \@firstoftwo
 \else \expandafter \@secondoftwo
 \fi
}%
\providecommand \natexlab [1]{#1}%
\providecommand \enquote  [1]{``#1''}%
\providecommand \bibnamefont  [1]{#1}%
\providecommand \bibfnamefont [1]{#1}%
\providecommand \citenamefont [1]{#1}%
\providecommand \href@noop [0]{\@secondoftwo}%
\providecommand \href [0]{\begingroup \@sanitize@url \@href}%
\providecommand \@href[1]{\@@startlink{#1}\@@href}%
\providecommand \@@href[1]{\endgroup#1\@@endlink}%
\providecommand \@sanitize@url [0]{\catcode `\\12\catcode `\$12\catcode
  `\&12\catcode `\#12\catcode `\^12\catcode `\_12\catcode `\%12\relax}%
\providecommand \@@startlink[1]{}%
\providecommand \@@endlink[0]{}%
\providecommand \url  [0]{\begingroup\@sanitize@url \@url }%
\providecommand \@url [1]{\endgroup\@href {#1}{\urlprefix }}%
\providecommand \urlprefix  [0]{URL }%
\providecommand \Eprint [0]{\href }%
\providecommand \doibase [0]{https://doi.org/}%
\providecommand \selectlanguage [0]{\@gobble}%
\providecommand \bibinfo  [0]{\@secondoftwo}%
\providecommand \bibfield  [0]{\@secondoftwo}%
\providecommand \translation [1]{[#1]}%
\providecommand \BibitemOpen [0]{}%
\providecommand \bibitemStop [0]{}%
\providecommand \bibitemNoStop [0]{.\EOS\space}%
\providecommand \EOS [0]{\spacefactor3000\relax}%
\providecommand \BibitemShut  [1]{\csname bibitem#1\endcsname}%
\let\auto@bib@innerbib\@empty
\bibitem [{\citenamefont {Ohtomo}\ and\ \citenamefont
  {Hwang}(2004)}]{ohtomo2004high}%
  \BibitemOpen
  \bibfield  {author} {\bibinfo {author} {\bibfnamefont {A.}~\bibnamefont
  {Ohtomo}}\ and\ \bibinfo {author} {\bibfnamefont {H.}~\bibnamefont {Hwang}},\
  }\bibfield  {title} {\bibinfo {title} {{A high-mobility electron gas at the
  LaAlO$_3$/SrTiO$_3$ heterointerface}},\ }\href@noop {} {\bibfield  {journal}
  {\bibinfo  {journal} {Nature}\ }\textbf {\bibinfo {volume} {427}},\ \bibinfo
  {pages} {423} (\bibinfo {year} {2004})}\BibitemShut {NoStop}%
\bibitem [{\citenamefont {Caviglia}\ \emph {et~al.}(2008)\citenamefont
  {Caviglia}, \citenamefont {Gariglio}, \citenamefont {Reyren}, \citenamefont
  {Jaccard}, \citenamefont {Schneider}, \citenamefont {Gabay}, \citenamefont
  {Thiel}, \citenamefont {Hammerl}, \citenamefont {Mannhart},\ and\
  \citenamefont {Triscone}}]{caviglia2008electric}%
  \BibitemOpen
  \bibfield  {author} {\bibinfo {author} {\bibfnamefont {A.}~\bibnamefont
  {Caviglia}}, \bibinfo {author} {\bibfnamefont {S.}~\bibnamefont {Gariglio}},
  \bibinfo {author} {\bibfnamefont {N.}~\bibnamefont {Reyren}}, \bibinfo
  {author} {\bibfnamefont {D.}~\bibnamefont {Jaccard}}, \bibinfo {author}
  {\bibfnamefont {T.}~\bibnamefont {Schneider}}, \bibinfo {author}
  {\bibfnamefont {M.}~\bibnamefont {Gabay}}, \bibinfo {author} {\bibfnamefont
  {S.}~\bibnamefont {Thiel}}, \bibinfo {author} {\bibfnamefont
  {G.}~\bibnamefont {Hammerl}}, \bibinfo {author} {\bibfnamefont
  {J.}~\bibnamefont {Mannhart}},\ and\ \bibinfo {author} {\bibfnamefont
  {J.-M.}\ \bibnamefont {Triscone}},\ }\bibfield  {title} {\bibinfo {title}
  {{Electric field control of the LaAlO$_3$/SrTiO$_3$ interface ground
  state}},\ }\href@noop {} {\bibfield  {journal} {\bibinfo  {journal} {Nature}\
  }\textbf {\bibinfo {volume} {456}},\ \bibinfo {pages} {624} (\bibinfo {year}
  {2008})}\BibitemShut {NoStop}%
\bibitem [{\citenamefont {Reyren}\ \emph {et~al.}(2007)\citenamefont {Reyren},
  \citenamefont {Thiel}, \citenamefont {Caviglia}, \citenamefont {Kourkoutis},
  \citenamefont {Hammerl}, \citenamefont {Richter}, \citenamefont {Schneider},
  \citenamefont {Kopp}, \citenamefont {R{\"u}etschi}, \citenamefont {Jaccard}
  \emph {et~al.}}]{reyren2007superconducting}%
  \BibitemOpen
  \bibfield  {author} {\bibinfo {author} {\bibfnamefont {N.}~\bibnamefont
  {Reyren}}, \bibinfo {author} {\bibfnamefont {S.}~\bibnamefont {Thiel}},
  \bibinfo {author} {\bibfnamefont {A.}~\bibnamefont {Caviglia}}, \bibinfo
  {author} {\bibfnamefont {L.~F.}\ \bibnamefont {Kourkoutis}}, \bibinfo
  {author} {\bibfnamefont {G.}~\bibnamefont {Hammerl}}, \bibinfo {author}
  {\bibfnamefont {C.}~\bibnamefont {Richter}}, \bibinfo {author} {\bibfnamefont
  {C.~W.}\ \bibnamefont {Schneider}}, \bibinfo {author} {\bibfnamefont
  {T.}~\bibnamefont {Kopp}}, \bibinfo {author} {\bibfnamefont {A.-S.}\
  \bibnamefont {R{\"u}etschi}}, \bibinfo {author} {\bibfnamefont
  {D.}~\bibnamefont {Jaccard}}, \emph {et~al.},\ }\bibfield  {title} {\bibinfo
  {title} {Superconducting interfaces between insulating oxides},\ }\href@noop
  {} {\bibfield  {journal} {\bibinfo  {journal} {Science}\ }\textbf {\bibinfo
  {volume} {317}},\ \bibinfo {pages} {1196} (\bibinfo {year}
  {2007})}\BibitemShut {NoStop}%
\bibitem [{\citenamefont {Perroni}\ \emph {et~al.}(2019)\citenamefont
  {Perroni}, \citenamefont {Cataudella}, \citenamefont {Salluzzo},
  \citenamefont {Cuoco},\ and\ \citenamefont {Citro}}]{perroni1}%
  \BibitemOpen
  \bibfield  {author} {\bibinfo {author} {\bibfnamefont {C.~A.}\ \bibnamefont
  {Perroni}}, \bibinfo {author} {\bibfnamefont {V.}~\bibnamefont {Cataudella}},
  \bibinfo {author} {\bibfnamefont {M.}~\bibnamefont {Salluzzo}}, \bibinfo
  {author} {\bibfnamefont {M.}~\bibnamefont {Cuoco}},\ and\ \bibinfo {author}
  {\bibfnamefont {R.}~\bibnamefont {Citro}},\ }\bibfield  {title} {\bibinfo
  {title} {{Evolution of topological superconductivity by orbital-selective
  confinement in oxide nanowires}},\ }\href
  {https://doi.org/10.1103/PhysRevB.100.094526} {\bibfield  {journal} {\bibinfo
   {journal} {Phys. Rev. B}\ }\textbf {\bibinfo {volume} {100}},\ \bibinfo
  {pages} {094526} (\bibinfo {year} {2019})}\BibitemShut {NoStop}%
\bibitem [{\citenamefont {Maiellaro}\ \emph {et~al.}(2019)\citenamefont
  {Maiellaro}, \citenamefont {Romeo}, \citenamefont {Perroni}, \citenamefont
  {Cataudella},\ and\ \citenamefont {Citro}}]{maiellaro2019unveiling}%
  \BibitemOpen
  \bibfield  {author} {\bibinfo {author} {\bibfnamefont {A.}~\bibnamefont
  {Maiellaro}}, \bibinfo {author} {\bibfnamefont {F.}~\bibnamefont {Romeo}},
  \bibinfo {author} {\bibfnamefont {C.~A.}\ \bibnamefont {Perroni}}, \bibinfo
  {author} {\bibfnamefont {V.}~\bibnamefont {Cataudella}},\ and\ \bibinfo
  {author} {\bibfnamefont {R.}~\bibnamefont {Citro}},\ }\bibfield  {title}
  {\bibinfo {title} {Unveiling signatures of topological phases in open kitaev
  chains and ladders},\ }\href@noop {} {\bibfield  {journal} {\bibinfo
  {journal} {Nanomaterials}\ }\textbf {\bibinfo {volume} {9}},\ \bibinfo
  {pages} {894} (\bibinfo {year} {2019})}\BibitemShut {NoStop}%
\bibitem [{\citenamefont {Barthelemy}\ \emph {et~al.}(2021)\citenamefont
  {Barthelemy}, \citenamefont {Bergeal}, \citenamefont {Bibes}, \citenamefont
  {Caviglia}, \citenamefont {Citro}, \citenamefont {Cuoco}, \citenamefont
  {Kalaboukhov}, \citenamefont {Kalisky}, \citenamefont {Perroni},
  \citenamefont {Santamaria} \emph {et~al.}}]{barthelemy2021quasi}%
  \BibitemOpen
  \bibfield  {author} {\bibinfo {author} {\bibfnamefont {A.}~\bibnamefont
  {Barthelemy}}, \bibinfo {author} {\bibfnamefont {N.}~\bibnamefont {Bergeal}},
  \bibinfo {author} {\bibfnamefont {M.}~\bibnamefont {Bibes}}, \bibinfo
  {author} {\bibfnamefont {A.}~\bibnamefont {Caviglia}}, \bibinfo {author}
  {\bibfnamefont {R.}~\bibnamefont {Citro}}, \bibinfo {author} {\bibfnamefont
  {M.}~\bibnamefont {Cuoco}}, \bibinfo {author} {\bibfnamefont
  {A.}~\bibnamefont {Kalaboukhov}}, \bibinfo {author} {\bibfnamefont
  {B.}~\bibnamefont {Kalisky}}, \bibinfo {author} {\bibfnamefont {C.~A.}\
  \bibnamefont {Perroni}}, \bibinfo {author} {\bibfnamefont {J.}~\bibnamefont
  {Santamaria}}, \emph {et~al.},\ }\bibfield  {title} {\bibinfo {title}
  {Quasi-two-dimensional electron gas at the oxide interfaces for topological
  quantum physics},\ }\href@noop {} {\bibfield  {journal} {\bibinfo  {journal}
  {Europhysics Letters}\ }\textbf {\bibinfo {volume} {133}},\ \bibinfo {pages}
  {17001} (\bibinfo {year} {2021})}\BibitemShut {NoStop}%
\bibitem [{\citenamefont {Davis}\ \emph {et~al.}(2018)\citenamefont {Davis},
  \citenamefont {Huang}, \citenamefont {Han}, \citenamefont {Venkatesan},
  \citenamefont {Chandrasekhar} \emph {et~al.}}]{davis2018anisotropic}%
  \BibitemOpen
  \bibfield  {author} {\bibinfo {author} {\bibfnamefont {S.}~\bibnamefont
  {Davis}}, \bibinfo {author} {\bibfnamefont {Z.}~\bibnamefont {Huang}},
  \bibinfo {author} {\bibfnamefont {K.}~\bibnamefont {Han}}, \bibinfo {author}
  {\bibfnamefont {T.}~\bibnamefont {Venkatesan}}, \bibinfo {author}
  {\bibfnamefont {V.}~\bibnamefont {Chandrasekhar}}, \emph {et~al.},\
  }\bibfield  {title} {\bibinfo {title} {Anisotropic superconductivity and
  frozen electronic states at the (111) laalo 3/srtio 3 interface},\
  }\href@noop {} {\bibfield  {journal} {\bibinfo  {journal} {Physical Review
  B}\ }\textbf {\bibinfo {volume} {98}},\ \bibinfo {pages} {024504} (\bibinfo
  {year} {2018})}\BibitemShut {NoStop}%
\bibitem [{\citenamefont {Monteiro}\ \emph {et~al.}(2019)\citenamefont
  {Monteiro}, \citenamefont {Vivek}, \citenamefont {Groenendijk}, \citenamefont
  {Bruneel}, \citenamefont {Leermakers}, \citenamefont {Zeitler}, \citenamefont
  {Gabay},\ and\ \citenamefont {Caviglia}}]{monteiro2019band}%
  \BibitemOpen
  \bibfield  {author} {\bibinfo {author} {\bibfnamefont {A.}~\bibnamefont
  {Monteiro}}, \bibinfo {author} {\bibfnamefont {M.}~\bibnamefont {Vivek}},
  \bibinfo {author} {\bibfnamefont {D.}~\bibnamefont {Groenendijk}}, \bibinfo
  {author} {\bibfnamefont {P.}~\bibnamefont {Bruneel}}, \bibinfo {author}
  {\bibfnamefont {I.}~\bibnamefont {Leermakers}}, \bibinfo {author}
  {\bibfnamefont {U.}~\bibnamefont {Zeitler}}, \bibinfo {author} {\bibfnamefont
  {M.}~\bibnamefont {Gabay}},\ and\ \bibinfo {author} {\bibfnamefont
  {A.}~\bibnamefont {Caviglia}},\ }\bibfield  {title} {\bibinfo {title} {{Band
  inversion driven by electronic correlations at the (111) LaAlO$_3$/SrTiO$_3$
  interface}},\ }\href@noop {} {\bibfield  {journal} {\bibinfo  {journal}
  {Physical Review B}\ }\textbf {\bibinfo {volume} {99}},\ \bibinfo {pages}
  {201102} (\bibinfo {year} {2019})}\BibitemShut {NoStop}%
\bibitem [{\citenamefont {Trama}\ \emph {et~al.}(2021)\citenamefont {Trama},
  \citenamefont {Cataudella},\ and\ \citenamefont
  {Perroni}}]{trama2021straininduced}%
  \BibitemOpen
  \bibfield  {author} {\bibinfo {author} {\bibfnamefont {M.}~\bibnamefont
  {Trama}}, \bibinfo {author} {\bibfnamefont {V.}~\bibnamefont {Cataudella}},\
  and\ \bibinfo {author} {\bibfnamefont {C.~A.}\ \bibnamefont {Perroni}},\
  }\bibfield  {title} {\bibinfo {title} {{Strain-induced topological phase
  transition at (111) ${\mathrm{SrTiO}}_{3}$-based heterostructures}},\ }\href
  {https://doi.org/10.1103/PhysRevResearch.3.043038} {\bibfield  {journal}
  {\bibinfo  {journal} {Phys. Rev. Research}\ }\textbf {\bibinfo {volume}
  {3}},\ \bibinfo {pages} {043038} (\bibinfo {year} {2021})}\BibitemShut
  {NoStop}%
\bibitem [{\citenamefont {Trama}\ \emph
  {et~al.}(2022{\natexlab{a}})\citenamefont {Trama}, \citenamefont
  {Cataudella}, \citenamefont {Perroni}, \citenamefont {Romeo},\ and\
  \citenamefont {Citro}}]{trama2022gate}%
  \BibitemOpen
  \bibfield  {author} {\bibinfo {author} {\bibfnamefont {M.}~\bibnamefont
  {Trama}}, \bibinfo {author} {\bibfnamefont {V.}~\bibnamefont {Cataudella}},
  \bibinfo {author} {\bibfnamefont {C.~A.}\ \bibnamefont {Perroni}}, \bibinfo
  {author} {\bibfnamefont {F.}~\bibnamefont {Romeo}},\ and\ \bibinfo {author}
  {\bibfnamefont {R.}~\bibnamefont {Citro}},\ }\bibfield  {title} {\bibinfo
  {title} {Gate tunable anomalous hall effect: Berry curvature probe at oxides
  interfaces},\ }\href@noop {} {\bibfield  {journal} {\bibinfo  {journal}
  {Physical Review B}\ }\textbf {\bibinfo {volume} {106}},\ \bibinfo {pages}
  {075430} (\bibinfo {year} {2022}{\natexlab{a}})}\BibitemShut {NoStop}%
\bibitem [{\citenamefont {Trama}\ \emph
  {et~al.}(2022{\natexlab{b}})\citenamefont {Trama}, \citenamefont
  {Cataudella}, \citenamefont {Perroni}, \citenamefont {Romeo},\ and\
  \citenamefont {Citro}}]{trama2022tunable}%
  \BibitemOpen
  \bibfield  {author} {\bibinfo {author} {\bibfnamefont {M.}~\bibnamefont
  {Trama}}, \bibinfo {author} {\bibfnamefont {V.}~\bibnamefont {Cataudella}},
  \bibinfo {author} {\bibfnamefont {C.~A.}\ \bibnamefont {Perroni}}, \bibinfo
  {author} {\bibfnamefont {F.}~\bibnamefont {Romeo}},\ and\ \bibinfo {author}
  {\bibfnamefont {R.}~\bibnamefont {Citro}},\ }\bibfield  {title} {\bibinfo
  {title} {{Tunable Spin and Orbital Edelstein Effect at (111)
  LaAlO$_3$/SrTiO$_3$ Interface}},\ }\href@noop {} {\bibfield  {journal}
  {\bibinfo  {journal} {Nanomaterials}\ }\textbf {\bibinfo {volume} {12}},\
  \bibinfo {pages} {2494} (\bibinfo {year} {2022}{\natexlab{b}})}\BibitemShut
  {NoStop}%
\bibitem [{\citenamefont {Walker}\ \emph {et~al.}(2014)\citenamefont {Walker},
  \citenamefont {De~La~Torre}, \citenamefont {Bruno}, \citenamefont {Tamai},
  \citenamefont {Kim}, \citenamefont {Hoesch}, \citenamefont {Shi},
  \citenamefont {Bahramy}, \citenamefont {King},\ and\ \citenamefont
  {Baumberger}}]{walker2014control}%
  \BibitemOpen
  \bibfield  {author} {\bibinfo {author} {\bibfnamefont {S.~M.}\ \bibnamefont
  {Walker}}, \bibinfo {author} {\bibfnamefont {A.}~\bibnamefont {De~La~Torre}},
  \bibinfo {author} {\bibfnamefont {F.~Y.}\ \bibnamefont {Bruno}}, \bibinfo
  {author} {\bibfnamefont {A.}~\bibnamefont {Tamai}}, \bibinfo {author}
  {\bibfnamefont {T.}~\bibnamefont {Kim}}, \bibinfo {author} {\bibfnamefont
  {M.}~\bibnamefont {Hoesch}}, \bibinfo {author} {\bibfnamefont
  {M.}~\bibnamefont {Shi}}, \bibinfo {author} {\bibfnamefont {M.}~\bibnamefont
  {Bahramy}}, \bibinfo {author} {\bibfnamefont {P.}~\bibnamefont {King}},\ and\
  \bibinfo {author} {\bibfnamefont {F.}~\bibnamefont {Baumberger}},\ }\bibfield
   {title} {\bibinfo {title} {{Control of a two-dimensional electron gas on
  SrTiO$_3$ (111) by atomic oxygen}},\ }\href@noop {} {\bibfield  {journal}
  {\bibinfo  {journal} {Physical Review Letters}\ }\textbf {\bibinfo {volume}
  {113}},\ \bibinfo {pages} {177601} (\bibinfo {year} {2014})}\BibitemShut
  {NoStop}%
\bibitem [{\citenamefont {King}\ \emph {et~al.}(2014)\citenamefont {King},
  \citenamefont {Mckeown~Walker}, \citenamefont {Tamai}, \citenamefont
  {De~La~Torre}, \citenamefont {Eknapakul}, \citenamefont {Buaphet},
  \citenamefont {Mo}, \citenamefont {Meevasana}, \citenamefont {Bahramy},\ and\
  \citenamefont {Baumberger}}]{king2014quasiparticle}%
  \BibitemOpen
  \bibfield  {author} {\bibinfo {author} {\bibfnamefont {P.}~\bibnamefont
  {King}}, \bibinfo {author} {\bibfnamefont {S.}~\bibnamefont
  {Mckeown~Walker}}, \bibinfo {author} {\bibfnamefont {A.}~\bibnamefont
  {Tamai}}, \bibinfo {author} {\bibfnamefont {A.}~\bibnamefont {De~La~Torre}},
  \bibinfo {author} {\bibfnamefont {T.}~\bibnamefont {Eknapakul}}, \bibinfo
  {author} {\bibfnamefont {P.}~\bibnamefont {Buaphet}}, \bibinfo {author}
  {\bibfnamefont {S.-K.}\ \bibnamefont {Mo}}, \bibinfo {author} {\bibfnamefont
  {W.}~\bibnamefont {Meevasana}}, \bibinfo {author} {\bibfnamefont
  {M.}~\bibnamefont {Bahramy}},\ and\ \bibinfo {author} {\bibfnamefont
  {F.}~\bibnamefont {Baumberger}},\ }\bibfield  {title} {\bibinfo {title}
  {Quasiparticle dynamics and spin--orbital texture of the srtio3
  two-dimensional electron gas},\ }\href@noop {} {\bibfield  {journal}
  {\bibinfo  {journal} {Nature communications}\ }\textbf {\bibinfo {volume}
  {5}},\ \bibinfo {pages} {1} (\bibinfo {year} {2014})}\BibitemShut {NoStop}%
\bibitem [{\citenamefont {Bahramy}\ \emph {et~al.}(2012)\citenamefont
  {Bahramy}, \citenamefont {King}, \citenamefont {De~La~Torre}, \citenamefont
  {Chang}, \citenamefont {Shi}, \citenamefont {Patthey}, \citenamefont
  {Balakrishnan}, \citenamefont {Hofmann}, \citenamefont {Arita}, \citenamefont
  {Nagaosa} \emph {et~al.}}]{bahramy2012emergent}%
  \BibitemOpen
  \bibfield  {author} {\bibinfo {author} {\bibfnamefont {M.}~\bibnamefont
  {Bahramy}}, \bibinfo {author} {\bibfnamefont {P.}~\bibnamefont {King}},
  \bibinfo {author} {\bibfnamefont {A.}~\bibnamefont {De~La~Torre}}, \bibinfo
  {author} {\bibfnamefont {J.}~\bibnamefont {Chang}}, \bibinfo {author}
  {\bibfnamefont {M.}~\bibnamefont {Shi}}, \bibinfo {author} {\bibfnamefont
  {L.}~\bibnamefont {Patthey}}, \bibinfo {author} {\bibfnamefont
  {G.}~\bibnamefont {Balakrishnan}}, \bibinfo {author} {\bibfnamefont
  {P.}~\bibnamefont {Hofmann}}, \bibinfo {author} {\bibfnamefont
  {R.}~\bibnamefont {Arita}}, \bibinfo {author} {\bibfnamefont
  {N.}~\bibnamefont {Nagaosa}}, \emph {et~al.},\ }\bibfield  {title} {\bibinfo
  {title} {Emergent quantum confinement at topological insulator surfaces},\
  }\href@noop {} {\bibfield  {journal} {\bibinfo  {journal} {Nature
  communications}\ }\textbf {\bibinfo {volume} {3}},\ \bibinfo {pages} {1}
  (\bibinfo {year} {2012})}\BibitemShut {NoStop}%
\bibitem [{\citenamefont {K.}\ and\ \citenamefont {F.}(2011)}]{Momma11Vesta}%
  \BibitemOpen
  \bibfield  {author} {\bibinfo {author} {\bibfnamefont {M.}~\bibnamefont
  {K.}}\ and\ \bibinfo {author} {\bibfnamefont {I.}~\bibnamefont {F.}},\
  }\bibfield  {title} {\bibinfo {title} {Vesta 3 for three-dimensional
  visualization of crystal, volumetric and morphology data},\ }\href@noop {}
  {\bibfield  {journal} {\bibinfo  {journal} {J. Appl. Crystallogr.}\ }\textbf
  {\bibinfo {volume} {44}},\ \bibinfo {pages} {1272} (\bibinfo {year}
  {2011})}\BibitemShut {NoStop}%
\bibitem [{\citenamefont {Pai}\ \emph {et~al.}(2018)\citenamefont {Pai},
  \citenamefont {Tylan-Tyler}, \citenamefont {Irvin},\ and\ \citenamefont
  {Levy}}]{pai2018physics}%
  \BibitemOpen
  \bibfield  {author} {\bibinfo {author} {\bibfnamefont {Y.-Y.}\ \bibnamefont
  {Pai}}, \bibinfo {author} {\bibfnamefont {A.}~\bibnamefont {Tylan-Tyler}},
  \bibinfo {author} {\bibfnamefont {P.}~\bibnamefont {Irvin}},\ and\ \bibinfo
  {author} {\bibfnamefont {J.}~\bibnamefont {Levy}},\ }\bibfield  {title}
  {\bibinfo {title} {{Physics of SrTiO$_3$-based heterostructures and
  nanostructures: a review}},\ }\href@noop {} {\bibfield  {journal} {\bibinfo
  {journal} {Reports on Progress in Physics}\ }\textbf {\bibinfo {volume}
  {81}},\ \bibinfo {pages} {036503} (\bibinfo {year} {2018})}\BibitemShut
  {NoStop}%
\bibitem [{\citenamefont {Xiao}\ \emph {et~al.}(2011)\citenamefont {Xiao},
  \citenamefont {Zhu}, \citenamefont {Ran}, \citenamefont {Nagaosa},\ and\
  \citenamefont {Okamoto}}]{xiao2011interface}%
  \BibitemOpen
  \bibfield  {author} {\bibinfo {author} {\bibfnamefont {D.}~\bibnamefont
  {Xiao}}, \bibinfo {author} {\bibfnamefont {W.}~\bibnamefont {Zhu}}, \bibinfo
  {author} {\bibfnamefont {Y.}~\bibnamefont {Ran}}, \bibinfo {author}
  {\bibfnamefont {N.}~\bibnamefont {Nagaosa}},\ and\ \bibinfo {author}
  {\bibfnamefont {S.}~\bibnamefont {Okamoto}},\ }\bibfield  {title} {\bibinfo
  {title} {Interface engineering of quantum hall effects in digital transition
  metal oxide heterostructures},\ }\href@noop {} {\bibfield  {journal}
  {\bibinfo  {journal} {Nature Communications}\ }\textbf {\bibinfo {volume}
  {2}},\ \bibinfo {pages} {1} (\bibinfo {year} {2011})}\BibitemShut {NoStop}%
\bibitem [{\citenamefont {Bruneel}(2020)}]{bruneel2020electronic}%
  \BibitemOpen
  \bibfield  {author} {\bibinfo {author} {\bibfnamefont {P.}~\bibnamefont
  {Bruneel}},\ }\emph {\bibinfo {title} {Electronic and spintronic properties
  of the interfaces between transition metal oxides}},\ \href@noop {} {Ph.D.
  thesis},\ \bibinfo  {school} {Universit{\'e} Paris-Saclay} (\bibinfo {year}
  {2020})\BibitemShut {NoStop}%
\bibitem [{\citenamefont {Neville}\ \emph {et~al.}(1972)\citenamefont
  {Neville}, \citenamefont {Hoeneisen},\ and\ \citenamefont
  {Mead}}]{neville1972permittivity}%
  \BibitemOpen
  \bibfield  {author} {\bibinfo {author} {\bibfnamefont {R.}~\bibnamefont
  {Neville}}, \bibinfo {author} {\bibfnamefont {B.}~\bibnamefont {Hoeneisen}},\
  and\ \bibinfo {author} {\bibfnamefont {C.}~\bibnamefont {Mead}},\ }\bibfield
  {title} {\bibinfo {title} {Permittivity of strontium titanate},\ }\href@noop
  {} {\bibfield  {journal} {\bibinfo  {journal} {Journal of Applied Physics}\
  }\textbf {\bibinfo {volume} {43}},\ \bibinfo {pages} {2124} (\bibinfo {year}
  {1972})}\BibitemShut {NoStop}%
\bibitem [{\citenamefont {Dunitz}\ and\ \citenamefont
  {Orgel}(1957)}]{dunitz1957electronic}%
  \BibitemOpen
  \bibfield  {author} {\bibinfo {author} {\bibfnamefont {J.}~\bibnamefont
  {Dunitz}}\ and\ \bibinfo {author} {\bibfnamefont {L.}~\bibnamefont {Orgel}},\
  }\bibfield  {title} {\bibinfo {title} {Electronic properties of
  transition-metal oxides-ii: cation distribution amongst octahedral and
  tetrahedral sites},\ }\href@noop {} {\bibfield  {journal} {\bibinfo
  {journal} {Journal of Physics and Chemistry of Solids}\ }\textbf {\bibinfo
  {volume} {3}},\ \bibinfo {pages} {318} (\bibinfo {year} {1957})}\BibitemShut
  {NoStop}%
\bibitem [{\citenamefont {Barrett}(1952)}]{barret52}%
  \BibitemOpen
  \bibfield  {author} {\bibinfo {author} {\bibfnamefont {J.~H.}\ \bibnamefont
  {Barrett}},\ }\bibfield  {title} {\bibinfo {title} {Dielectric constant in
  perovskite type crystals},\ }\href {https://doi.org/10.1103/PhysRev.86.118}
  {\bibfield  {journal} {\bibinfo  {journal} {Phys. Rev.}\ }\textbf {\bibinfo
  {volume} {86}},\ \bibinfo {pages} {118} (\bibinfo {year} {1952})}\BibitemShut
  {NoStop}%
\bibitem [{\citenamefont {Delugas}\ \emph {et~al.}(2011)\citenamefont
  {Delugas}, \citenamefont {Filippetti}, \citenamefont {Fiorentini},
  \citenamefont {Bilc}, \citenamefont {Fontaine},\ and\ \citenamefont
  {Ghosez}}]{delugas2011spontaneous}%
  \BibitemOpen
  \bibfield  {author} {\bibinfo {author} {\bibfnamefont {P.}~\bibnamefont
  {Delugas}}, \bibinfo {author} {\bibfnamefont {A.}~\bibnamefont {Filippetti}},
  \bibinfo {author} {\bibfnamefont {V.}~\bibnamefont {Fiorentini}}, \bibinfo
  {author} {\bibfnamefont {D.~I.}\ \bibnamefont {Bilc}}, \bibinfo {author}
  {\bibfnamefont {D.}~\bibnamefont {Fontaine}},\ and\ \bibinfo {author}
  {\bibfnamefont {P.}~\bibnamefont {Ghosez}},\ }\bibfield  {title} {\bibinfo
  {title} {Spontaneous 2-dimensional carrier confinement at the n-type
  srtio3/laalo3 interface},\ }\href@noop {} {\bibfield  {journal} {\bibinfo
  {journal} {Physical Review Letters}\ }\textbf {\bibinfo {volume} {106}},\
  \bibinfo {pages} {166807} (\bibinfo {year} {2011})}\BibitemShut {NoStop}%
\bibitem [{\citenamefont {Khomskii}(2014)}]{khomskii2014transition}%
  \BibitemOpen
  \bibfield  {author} {\bibinfo {author} {\bibfnamefont {D.}~\bibnamefont
  {Khomskii}},\ }\href@noop {} {\emph {\bibinfo {title} {Transition metal
  compounds}}}\ (\bibinfo  {publisher} {Cambridge University Press},\ \bibinfo
  {year} {2014})\BibitemShut {NoStop}%
\bibitem [{\citenamefont {De~Luca}\ \emph {et~al.}(2018)\citenamefont
  {De~Luca}, \citenamefont {Di~Capua}, \citenamefont {Di~Gennaro},
  \citenamefont {Sambri}, \citenamefont {Granozio}, \citenamefont
  {Ghiringhelli}, \citenamefont {Betto}, \citenamefont {Piamonteze},
  \citenamefont {Brookes},\ and\ \citenamefont {Salluzzo}}]{de2018symmetry}%
  \BibitemOpen
  \bibfield  {author} {\bibinfo {author} {\bibfnamefont {G.}~\bibnamefont
  {De~Luca}}, \bibinfo {author} {\bibfnamefont {R.}~\bibnamefont {Di~Capua}},
  \bibinfo {author} {\bibfnamefont {E.}~\bibnamefont {Di~Gennaro}}, \bibinfo
  {author} {\bibfnamefont {A.}~\bibnamefont {Sambri}}, \bibinfo {author}
  {\bibfnamefont {F.~M.}\ \bibnamefont {Granozio}}, \bibinfo {author}
  {\bibfnamefont {G.}~\bibnamefont {Ghiringhelli}}, \bibinfo {author}
  {\bibfnamefont {D.}~\bibnamefont {Betto}}, \bibinfo {author} {\bibfnamefont
  {C.}~\bibnamefont {Piamonteze}}, \bibinfo {author} {\bibfnamefont
  {N.}~\bibnamefont {Brookes}},\ and\ \bibinfo {author} {\bibfnamefont
  {M.}~\bibnamefont {Salluzzo}},\ }\bibfield  {title} {\bibinfo {title}
  {{Symmetry breaking at the (111) interfaces of SrTiO$_3$ hosting a
  two-dimensional electron system}},\ }\href@noop {} {\bibfield  {journal}
  {\bibinfo  {journal} {Physical Review B}\ }\textbf {\bibinfo {volume} {98}},\
  \bibinfo {pages} {115143} (\bibinfo {year} {2018})}\BibitemShut {NoStop}%
\end{thebibliography}%


%

%


\end{document}